\documentclass[fleqn,usenatbib]{mnras}

\usepackage{natbib}

\usepackage[T1]{fontenc}
\usepackage{ae,aecompl}
\usepackage{verbatim}

\usepackage{ulem}
\usepackage[dvipsnames]{xcolor}

\usepackage{graphicx}	% Including figure files
\usepackage{amsmath}	% Advanced maths commands
\usepackage{amssymb}	% Extra maths symbols
\usepackage{newtxtext,newtxmath}

\title[Connecting steady and flare emission in Mrk\,421]{Connecting steady emission and Very High Energy flaring states in blazars: the case of Mrk\,421} 

\author[A. Dmytriiev]{
A. Dmytriiev,$^{1}$\thanks{E-mail: anton.dmytriiev@obspm.fr}
H. Sol $^{1}$
and A. Zech$^{1}$
\\
$^{1}$LUTH, Observatoire de Paris, Universit\'e de Paris, CNRS, PSL University, 5 Pl. Jules Janssen, 92195 Meudon, France\\
}

\date{Accepted XXX. Received YYY; in original form ZZZ}

\pubyear{2021}

\begin{document}
\label{firstpage}
\pagerange{\pageref{firstpage}--\pageref{lastpage}}
\maketitle

\begin{abstract}
Various attempts have been made in the literature at describing the origin and the physical mechanisms behind flaring events in blazars with radiative emission models, but detailed properties of multi-wavelength (MWL) light curves still remain difficult to reproduce. We have developed a versatile radiative code, based on a time-dependent treatment of particle acceleration, escape and radiative cooling, allowing us to test different scenarios to connect the continuous low-state emission self-consistently with that during flaring states. We consider flares as weak perturbations of the quiescent state and apply this description to the February 2010 MWL flare of Mrk\,421, the brightest Very High Energy (VHE) flare ever detected from this archetypal blazar, focusing on interpretations with a minimum number of free parameters. A general criterion is obtained, which disfavours a one-zone model connecting low and high state under our assumptions. A two-zone model combining physically connected acceleration and emission regions yields a satisfactory interpretation of the available time-dependent MWL light curves and spectra of Mrk\,421, although certain details remain difficult to reproduce. The two-zone scenario finally proposed for the complex quiescent and flaring VHE emitting region involves both Fermi-I and Fermi-II acceleration mechanisms, respectively at the origin of the quiescent and flaring emission.
\end{abstract}


%%%%%%%%%%%%%%%%%%%%%%%%%%%%%%%%%%%%%%%%%%%%%%%%%%

%%%%%%%%%%%%%%%%% BODY OF PAPER %%%%%%%%%%%%%%%%%%

\section{Introduction}

\subsection{Mrk\,421 at VHE $\gamma$-rays}

% a brief description of blazar characteristics, particularly BL Lacs, more particularly Mrk421

Markarian\,421 (Mrk\,421 , RA = 11$^{\text{h}}$ 04$^{\text{m}}$ 27$^{\text{s}}$ , Dec = +38$^{\circ}$ 12$^{\prime}$ 32$^{\prime \prime}$ , $z=0.031$) is generally the brightest extragalactic $\gamma$-ray source in the Very-High-Energy (VHE; energy range above about 100 GeV) $\gamma$-ray sky. It is also the most nearby representative of the blazar type of Active Galactic Nuclei (AGN), the sub-class of radio-loud AGN with jets aligned along the line of sight, and one of the best studied high-frequency-peaked BL Lac objects (HBLs). The flux from this type of sources is highly variable, with flux levels typically changing by more than one order of magnitude during flaring activity. 

To reveal the origin of the flaring behaviour of blazars, one needs to understand the changes of the physical conditions in the source between the quiescent and flaring states. This requires that timing and spectral properties of the source are well constrained from observations in these two states. In the VHE $\gamma$-ray band, where the fastest variability is observed, this is generally not the case. Imaging Atmospheric Cherenkov Telescopes (IACTs), which are the most sensitive observational tools in this energy range, have fields of view of a few degrees, so the continuous long-term monitoring of a large number of sources distributed across the sky is not possible. Despite the fast development of target of opportunity (ToO) observations, it remains difficult to promptly detect and observe flaring events, and the vast majority of the outbursts observed in VHE $\gamma$-rays are incomplete in time coverage. 
Moreover, MWL alerts and campaigns request the participation of many different ground-based and space instruments, which are complex to coordinate. In this context, Mrk\,421 is one of the most observed sources at VHE, providing particularly rich data sets.

The long-term variability of Mrk\,421 in the VHE band was studied first with the Whipple telescope over a 14-year period \citep{acciari2014}. Significant flux variations were found on time-scales ranging from a few minutes to years. Evidence was seen for correlations with the X-ray flux at monthly and yearly time-scales. A study of the source's long-term MWL behaviour from 2007 to 2009 including data from the MAGIC IACTs found VHE flux variations in the range from  $1.3 \times 10^{-11} < F(>400 \text{ GeV}) <  3.1 \times 10^{-10}$ ph cm$^{-2}$ s$^{-1}$ over this period \citep{ahnen2016}. During both low and high states the observed variability is higher in the high-energy bands than in the low-energy bands. A direct correlation of the VHE emission with the X-ray band was confirmed, while the authors found no significant correlation with the radio and optical bands.

Several flaring events from this source were studied for instance by \cite{caowang} (June 2008 flare), \cite{flaremagicmarch2010} (March 2010 flare), \cite{flares2012and2013} (2012 and 2013 flares), \cite{flaremagicapril2013} (April 2013 flare). To explain the observed correlated behaviour between X-rays and $\gamma$-rays during flux fluctuations, a synchrotron self-Compton (SSC) emission scenario is usually preferred.

% observations of the flare on February 2010
\subsection{February 2010 VHE flare of Mrk\,421}

We focus in this paper on the brightest VHE flare as of yet, reaching a level of about 27 Crab units above 1\,TeV, observed in February 2010 and followed with a range of instruments from the radio band up to VHE $\gamma$-rays as described by \cite{hessobsflare}, \cite{shukla2012}, \cite{singh2014}, \cite{veritasobsflare}. The flux of the source was seen to vary over seven days from the optical to the VHE band, with a peak on MJD\,55243 or MJD\,55244 and a possible time delay of the low-energy signal relatively to the VHE $\gamma$-rays. The X-ray flux increased by more than a factor of two in one day and spectral hardening with increasing flux was observed in X-rays and $\gamma$-rays. The detailed analysis of MWL variability and correlation studies recently published by the VERITAS and MAGIC collaboration with their MWL partners \citep{veritasobsflare} reports short time-scale variability with the emission rapidly varying during the main flare on the 10 min time-scale, and a complex VHE versus X-ray flux relationship. We will focus here on the MWL behaviour of the flare emission at a daily time-scale and do not try to directly reproduce very rapid variability.

\subsection{Physical modelling of blazar flares}

To explain the emergence of flares within relativistic jets, one generally distinguishes two possible types of scenarios. In the first type, variations of macrophysical properties of the emitting region, like its global geometry and kinematics, are responsible for launching outbursts. For instance, the relativistic Doppler factor can be increased by a change of viewing angle or bulk Lorentz factor, which can lead to stronger emission boosting and launch flares (e.g.\ \cite{casadio} ; \cite{larionov2016} ; \cite{raiteri}). This happens if the emitting region is moving along a curved trajectory, due to jet bending \citep{abdonature2010} or its helical configuration \citep{villataraiteri1999}. In the second type, the observed flux variability is considered to be due to the microphysics inside the VHE emitting zone and the subsequent evolution of the population of radiating particles caused by various physical processes, for example enhanced injection (e.g.\ \cite{mk1997}), and different particle acceleration mechanisms due to the development of shocks (e.g.\ \cite{marschergear1985} ; \cite{sikora2001} ; \cite{bottcherbaring}), turbulence (e.g.\  \cite{boutelier2008} ; \cite{tammiduffy} ; \cite{tramacere} ; \cite{shuklaetal2016}) or magnetic reconnection (e.g.\ \cite{giannios} ; \cite{dalpino2010} ; \cite{shuklaetal2018} ; \cite{shuklamannheim2020}). Emission models often combine the two types of mechanisms in order to reproduce extreme flares (e.g.\ \cite{marscher2014} ; \cite{katarzynskistat}).

\subsection{Previous interpretations of the February 2010 flare}

% interpretations of the February 2010 flare
A first interpretation of the remarkable 2010 flare of Mrk\,421 using spectral modelling is given by \cite{shukla2012}, \cite{singh2014} and \cite{singh2017}. \cite{singh2014} model the daily averaged spectral energy distribution (SED) on MJD\,55243 with a standard one-zone SSC scenario, assuming an instantaneous electron distribution that follows a broken power law. The variability time-scale is used to constrain the source parameters relying on the usual light travel time argument. 

A similar approach by \cite{shukla2012} provides instantaneous models for several SED data sets before, during and after the flare. The authors distinguish between a peak in X-rays and soft $\gamma$-rays on MJD\,55243 and a peak possibly delayed in the VHE emission on MJD\,55244. In their scenario, the flaring state arises from a change in several parameters, including Doppler factor, magnetic field strength, electron energy density and spectral index, which is attributed to electron acceleration in a strong shock. 

\cite{singh2017} propose a first model of the MWL light curve during the flare, which they find to be asymmetric in the high-energy bands, with a rise that is faster than the decay. They model the flux evolution detected by {\it Swift}-XRT, MAXI, {\it Fermi}-LAT and TACTIC with a one-zone model by adjusting a time-dependent injection function, assuming a constant spectral index of about 2.0. 

The most recent interpretation of the MWL spectral evolution during the flare is given by \cite{banerjee}, based on a time-dependent internal shock model \citep{joshibottcher}. Four spectral states before and during the flare were fit individually by varying key parameters of this multi-zone model. The authors find notably that the particle injection spectrum hardens during the flare, which they interpret as a shift from a dominant shock acceleration
mechanism during the low state to stochastic acceleration during the flaring state.

\subsection{The present model for the February 2010 flare}

In this paper we explore the approach based on the evolution of the particle distribution, assuming constant physical parameters of the emitting zone (such as its size, Doppler factor and magnetic field strength), and propose time-dependent SSC scenarios trying to connect self-consistently the long-term low-state emission from Mrk\,421 as described by \cite{abdo2011}, with the prominent February 2010 flare. To understand the nature of the outburst and the relevant physical processes involved, we perform detailed physical modelling of the data set, including stochastic and shock acceleration in addition to mere particle injection and produce fits of the MWL light curves in optical to VHE bands as well as of spectral measurements during different activity states, having in mind to propose a coherent global picture of the source with a physical model in which the variability pattern observed during the flare arises naturally from the quiescent state of the source due to a weak perturbation in or around the long-term emission region.

The data sets on Mrk\,421 studied in this work are presented in Section~\ref{sec:data}. In Section~\ref{sec:model}, we introduce the time-dependent emission code we have developed, which serves as a building block for the scenarios discussed in the following sections. To limit the number of free parameters, homogeneous one-zone models are explored first, and then extended to two zones when necessary to provide a satisfactory description of the data set. In Section~\ref{sec:scenario_ls} we analyse and reproduce the quiescent state of Mrk\,421. A criterion to test the validity of one-zone models is established in Section~\ref{sec:generalvalcrit} for individual SSC flares. Section~\ref{sec:scenario_flare} shows how a two-zone model can reproduce the MWL light curves of the flare. A general discussion of the results and perspectives is presented in Section~\ref{sec:discussion}.

\section{Observational data}
\label{sec:data}

\subsection{Data set for the low state}

\cite{abdo2011} present a composite MWL data set of Mrk\,421 in a low activity state, obtained as a result of a 4.5-month-long observational campaign on the source (19 January -- 1 June, 2009). Many instruments were participating in the campaign, including the VLBA, radio telescopes from the F-GAMMA program, optical and near-infrared telescopes from the GASP-WEBT program, {\it Swift}, RXTE, {\it Fermi}-LAT and MAGIC, among others. During this time period, the source was found in a relatively low flux state, and displayed almost no variability in all energy bands. The combined time-averaged measurement reported by the authors thus serves as a good proxy for a long-term quiescent or steady-state SED of Mrk\,421.

\subsection{Data set for the February 2010 flare}

A giant flare was observed during the period February 10 -- 23, 2010 (MJD\,55237 -- 55250). The source showed variability from the optical to the VHE $\gamma$-ray range. In X-rays and in the VHE band, the energy flux increased by a factor of about 5 to 10, reaching a peak around February 16, 2010 (MJD\,55243.5 -- 55244.0). The X-ray flux was found to correlate with the one in the TeV band \citep{shukla2012}. Variability was observed at two time-scales: $\sim$1 d (in all energy bands), and $\sim$ 1 h (intra-night variability) at TeV energies (\cite{veritasobsflare} ; \cite{shukla2012}). A secondary flaring event is observed after the main flare in X-rays at MJD\,55249, however it is not significant at MeV-GeV energies in the {\it Fermi} light curve. The flux increase in the optical {\it V}-band is very modest at around $20-30$ per cent and the variability in the radio band is negligible \citep{shukla2012}.

We have compiled the published light curves over all wavelength bands, as well as the available spectra. VHE observations of the outburst were performed with H.E.S.S.\ from MJD\,55245.0 to MJD\,55247.0 \citep{hessobsflare}, and VERITAS around MJD\,55244.3 (\cite{fortson} ; \cite{veritasobsflare}). VERITAS also monitored the source during the three following nights. Unfortunately, these instruments observed neither the rise of the flare nor the very peak, with VERITAS starting data-taking roughly one day after the estimated flux maximum, and H.E.S.S.\ -- around 1.5 d after the estimated peak. More complete time coverage in the VHE range, albeit with smaller sensitivity, was achieved by the HAGAR array \citep{shukla2012}, monitoring the source during February 13 -- 19, and by the TACTIC Cherenkov telescope \citep{singh2014}, observing during February 10 -- 23. The VHE flux recorded with the TACTIC telescope appeared significantly lower than the flux observed by the VERITAS and H.E.S.S.\ telescope arrays when compared in the same energy range and during the same time period, assuming the average spectral shape measured with H.E.S.S. We ascribe this discrepancy to uncertainties in the absolute calibration and rescale the TACTIC light curve by a constant factor of 5.7 to ensure consistency with VERITAS and H.E.S.S.\ fluxes.  

Data during the flare from the {\it Fermi}-LAT instrument in the MeV-GeV $\gamma$-ray band were published by \cite{singh2014} (who used instrument response functions P7SOURCE\_V6 in their analysis) and by \cite{veritasobsflare} (who used more recent instrument response functions P8R2\_SOURCE\_V6), X-ray data from {\it Swift}-XRT by \cite{shukla2012} and \cite{singh2014}, {\it Swift}-BAT data by \cite{shukla2012}, MAXI data by \cite{singh2014}, RXTE-PCA and RXTE-ASM data by \cite{shukla2012}. Optical data are available from the SPOL telescope and radio data from OVRO and were taken from \cite{shukla2012}.

To constrain the time-dependent SED, we compare our model results to spectra measured with XRT in soft X-ray and {\it Swift}-BAT in the hard X-ray band near the peak of the flare (February 16, 2010), to the {\it Fermi}-LAT uncertainty band for the spectrum (all taken from \cite{singh2014}), as well as to spectra from H.E.S.S.\ (time-averaged SED for the period February 17 -- 20, 2010) \citep{hessobsflare} and VERITAS (based on 5 h of data taken on February 17, 2010, roughly one day after the estimated peak of the flare) \citep{fortson}.

To compare the published count rates from the RXTE-ASM and {\it Swift}-BAT instruments to the light curves from the other wavelength bands, they were converted into energy fluxes first. The RXTE-ASM count rates were converted following the prescriptions by \cite{grimm} and \cite{chitnis} for a Crab-like spectrum. The difference between the Crab spectrum and an average spectrum of Mrk\,421 in this energy range was evaluated to add a systematic error of about 2 per cent to the uncertainty of the energy flux. It was also verified that photoabsorption on the Galactic hydrogen column is negligible in this energy range. 

For the conversion of the {\it Swift}-BAT count rates to a photon flux, the count rate at the peak is normalised with the spectrum presented by \cite{singh2014}, which we fit with a power law and integrate over the energy range of 15 to 50\,keV. A systematic uncertainty from the fitting procedure is added to the error on the resulting photon flux.

\section{The EMBLEM time-dependent blazar model}
\label{sec:model}
% Brief description of the EMBLEM model.

We adopt the conventional leptonic SSC scenario for the origin of blazar $\gamma$-ray emission, assuming it originates from a spherical region of radius $R_\text{b}$ (a `blob'), filled with a homogeneous relativistic electron-positron plasma embedded in a tangled magnetic field $B$ with a uniform flux density. The plasma blob is relativistically moving along the jet axis with a Doppler factor $\delta_\text{b}$. From now on we will refer to electrons and positrons as simply electrons. The electron population in the emitting zone is evolving due to several processes. Particles are injected into the blob with a spectrum $Q_{\text{inj}}(\gamma)$ which may depend on time, and may gain energy due to acceleration by shock (Fermi-I) or stochastic (Fermi-II) processes. The electrons confined in the blob radiate synchrotron and inverse Compton (IC) emission, and cool through radiative losses, comprising synchrotron and inverse Compton cooling. We also take into account synchrotron self-absorption. The particles escape the emitting zone at a characteristic time-scale $t_{\text{esc}}$. We assume that the size of the blob remains constant at first order, which allows to drastically reduce the number of free parameters. We also neglect adiabatic losses assuming that during the quiescent state the emitting zone is confined inside the jet (e.g.\ by its pressure external to the blob) with constant size, and internal $\gamma$-$\gamma$ absorption, which is typically negligible in HBLs (e.g.\ \cite{katarzynskistat}). Also, we disregard the emission from the extended jet, which we assume is only significant at very low energies. We treat the flaring behaviour as originating from the varying spectrum of the electron population in the blob.

The time evolution of the electron spectrum $N_\text{e}(\gamma,t)$ in the emitting zone is governed by a kinetic (Fokker-Planck) equation representing a continuity equation in phase space. Its general form, taking into account the above mentioned physical processes in the blob, for the case of `hard-sphere' turbulence (see sub-section~\ref{subsec:stochasticacc}) is (e.g.\ \cite{kardashev} ; \cite{tramacere}):

\begin{multline} \label{eq:kineticeqgeneral}
 		\dfrac{\partial N_\text{e}(\gamma,t)}{\partial t} = \dfrac{\partial}{\partial \gamma} \left[ (b_\text{c} \gamma^2 - a\gamma - 2D_0\gamma ) \cdot N_\text{e}(\gamma,t) \right] \, + \\ + \, \dfrac{\partial}{\partial \gamma}\left(D_0\gamma^2 \dfrac{\partial N_\text{e}(\gamma,t)}{\partial \gamma} \right) - \dfrac{N_\text{e}(\gamma,t)}{t_{\text{esc}}} + Q_{\text{inj}}(\gamma,t)
 		\end{multline}

The physical processes behind the different terms on the right hand side are described in the following subsections.

\subsection{Electron cooling}

The term $b_\text{c} \gamma^2$ corresponds to the total radiative cooling rate, comprising the synchrotron and the inverse Compton cooling rate: 

\begin{equation*}
    b_\text{c} \gamma^2 = |\dot{\gamma}_{\text{syn}}| + |\dot{\gamma}_{_{\text{IC}}}|
\end{equation*}

The synchrotron cooling rate is (e.g.\ \cite{chiaberge1999}):

\begin{equation*}
    |\dot{\gamma}_{\text{syn}}| = \frac{4 \sigma_\text{T}}{3 m_\text{e} c} \gamma^2 \, U_\text{B}
\end{equation*}

where $U_\text{B} = \frac{B^2}{2 \mu_0}$ is the magnetic energy density, $\sigma_\text{T}$ is the Thomson cross-section, $m_\text{e}$ is the electron rest mass, $c$ is the vacuum speed of light. 

The inverse Compton cooling rate is given by \cite{moderski2005}:

\begin{equation*}
    |\dot{\gamma}_{_{\text{IC}}}| = \frac{4 \sigma_\text{T}}{3 m_\text{e} c} \gamma^2 \, \int_{\epsilon^{\prime}_{\text{min}}}^{\epsilon^{\prime}_{\text{max}}} f_{_{\text{KN}}}(4\gamma \epsilon^{\prime}) \, u^{\prime}_{_{\text{syn}}}(\epsilon^{\prime}) \, d\epsilon^{\prime} 
\end{equation*}

where $\epsilon^{\prime} = \frac{h \nu^{\prime}}{m_\text{e} c^2}$ is the energy of seed (synchrotron) photons in the reference frame of the emitting zone in the units of the electron rest energy, $u^{\prime}_{_{\text{syn}}}(\epsilon^{\prime})$ is the distribution of the energy density of the synchrotron photons, and the function $f_{_{\text{KN}}}(x)$ is approximated as (including Klein-Nishina effects):

\begin{equation*}
    f_{_{\text{KN}}}(x) \simeq \begin{cases} (1+x)^{-1.5}, & \mbox{for } x < 10^4 \\ \frac{9}{2 x^2} \, (\text{ln}\,x - 11/6), & \mbox{for } x \geq 10^4 \end{cases}
\end{equation*}

The term $b_\text{c}$ in the total cooling rate is thus

\begin{equation}
    b_\text{c} = \frac{4 \sigma_\text{T}}{3 m_\text{e} c} \, \left[ \, U_\text{B} \, + \, \int_{\epsilon^{\prime}_{\text{min}}}^{\epsilon^{\prime}_{\text{max}}} f_{_{\text{KN}}}(4\gamma \epsilon^{\prime}) \, u^{\prime}_{_{\text{syn}}}(\epsilon^{\prime}) \, d\epsilon^{\prime} \, \right]
\end{equation}

and the characteristic cooling time of electrons is $t_{_{\text{cool}}}(\gamma) = \frac{1}{b_\text{c} \gamma}$.

It is worth to note, that the standard kinetic equation (Eq.~\ref{eq:kineticeqgeneral}) could be formally not applicable in the case where the inverse Compton cooling in Klein-Nishina regime becomes important, since the Eq.~\ref{eq:kineticeqgeneral} is derived assuming that the electrons lose only a small fraction of their energy in one interaction, while this is no longer the case in Klein-Nishina regime. However, inverse Compton cooling is usually negligible in HBLs, so that the standard kinetic equation can be applied to Mrk\,421.

\subsection{Shock acceleration}

Fermi-I (or diffusive shock) acceleration operates at the fronts of hydrodynamical shock waves, in the presence of velocity discontinuities. The principal mechanism of acceleration of relativistic charged particles by a strong shock is discussed by \cite{bell1978}. As a collisionless shock propagates through magnetised plasma, relativistic particles scatter on some turbulences or Alfv\'en waves \citep{wentzel} in the downstream and upstream regions, gradually gaining energy at each crossing of the shock front. The same scattering processes which entrap the particles near the shock are also responsible for their escape. The downstream plasma is receding from the shock front, leading to advection of the accelerated particles away from the shock, since their velocity distribution is isotropic in the medium frame. The spectrum of the particles escaping the shock follows a power law $dN_\text{e} / d\gamma \propto \gamma^{-\alpha_{\text{es}}}$ with an index $2 \leq \alpha_{\text{es}} \leq 2.5$, depending on the gas compression ratio.

Shock acceleration is a ubiquitous phenomenon thought to occur in many astrophysical systems, in particular in AGN jets (e.g.\ \cite{marschergear1985}). It is therefore natural to consider the possibility that the blazar outbursts could be triggered by a shock passing through the emitting zone. In addition, continuously operating on a long-term basis, the Fermi-I process could also serve as an efficient pre-acceleration mechanism supplying high-energy particles to the blob. Pre-accelerated particles can be further re-accelerated by another shock or by second-order Fermi mechanisms.  

We treat the shock acceleration process with the kinetic approach of Eq.~\ref{eq:kineticeqgeneral}. The Fermi-I process is considered as a systematic energy gain, and is described in the kinetic equation by the term $a\gamma$, which is the Fermi-I acceleration rate, a quantity proportional to the particle energy gain per unit of time: $\dot{\gamma}_{_{\text{FI}}} = a\gamma$. The characteristic time-scale of the shock acceleration is $t_{_{\text{FI}}} = 1 / a$. Throughout the paper, the $a$-term is put equal to zero when the shock acceleration is not active, and can be suddenly activated as needed to start the acceleration process. This introduces an additional important parameter $t_{_{\text{dur,FI}}}$ for the duration of the acceleration phase, namely $t_{_{\text{dur,FI}}} = t_{\text{cs}}$ for the crossing time or the lifetime of the shock defined in the source frame.

\subsection{Turbulence and stochastic acceleration}
\label{subsec:stochasticacc}

The presence of turbulence in the region can also lead to Fermi-II acceleration of particles. The energy is injected into the region at the largest spatial (stirring) scale, comparable to the size of the region, and cascades down to smaller scales, until the minimal one at which viscosity losses become dominant. The turbulent motion in the magnetised plasma produces a stochastic component of the magnetic field $\delta B$ in addition to the main (ordered) component $B_0$, which perturbs the plasma and excites Alfv\'en waves. In the quasi-linear framework that we consider here (e.g.\ \cite{schlickeiser1989}, \cite{jaekel1992}), the magnetohydrodynamic (MHD) turbulence is described by a combination of Alfv\'en waves with different wave numbers, forming a continuous wave spectrum. Particles of the plasma interact with the Alfv\'en waves and may exchange energy and momentum, leading to a gradual energy gain in a stochastic manner (see e.g.\ \cite{dermer}). The momentum diffusion coefficient describing momentum--energy gain by a particle is controlled by the wave-turbulence power spectrum, which has a form $W(k) \propto k^{-q}$, where $k = 2 \pi / \lambda$ is the wavenumber. The power spectrum is normalised as follows,  $\int_{k_{\text{min}}}^{k_{\text{max}}} W(k) \, dk = \frac{\delta B^2}{2 \mu_0}$ which is the total energy density stored in the magnetic fluctuations. The minimum and the maximum wavenumbers correspond to the longest ($\lambda_{\text{max}}$) and the shortest wavelength ($\lambda_{\text{min}}$) in the Alfv\'en spectrum accordingly. The spectral index $q = 3/2$ for the Kraichnan turbulence, $q = 5/3$ for the Kolmogorov turbulence, and $q = 2$ for the `hard-sphere' approximation we adopt here to describe the Fermi-II acceleration (see e.g.\ \cite{zhou} ; \cite{asano2018}). The `hard-sphere' turbulent spectrum is chosen as it favours the production of the brightest flares due to the most efficient re-acceleration of high-energy particles (e.g.\ \cite{becker2006}). The choice is additionally justified based on the observational properties of the Mrk\,421 February 2010 flare (cf. sub-section~\ref{subsec:twozonemodel}). 

The momentum diffusion coefficient for the process of stochastic acceleration in quasi-linear theory is given by (e.g.\ \cite{schlickeiser1989} ; \cite{osullivan}):

\begin{equation} \label{eq:momdiffcoefarbq}
   D_{_{p,\text{FII}}}(p) \, \approx \, \beta_\text{A}^2 \, \dfrac{\delta B^2}{B_{0}^2} \, \left( \dfrac{r_{_{\text{L}}}}{\lambda_{\text{max}}} \right)^{q-1} \, \dfrac{c p^2}{r_{_{\text{L}}}}   
\end{equation}

where $\beta_\text{A}$ is the Alfv\'en speed in the units of the speed of light, $r_{_{\text{L}}} = \frac{p}{e B_0}$ is the Larmor radius, and $p = \sqrt{\gamma^2 - 1} \, m_\text{e} c$ is the electron momentum. The quantity $\frac{\delta B^2}{B_{0}^2}$ is commonly referred to as `turbulence level'. The characteristic time-scale of Fermi-II acceleration process is then 

\begin{equation} \label{eq:tfermitwoarbq}
    t_{_{\text{FII}}} \, = \, \frac{p^2}{D_{_{p,\text{FII}}}(p)} \, = \, \dfrac{1}{\beta_\text{A}^2} \, \dfrac{B_0^2}{\delta B^2} \, \dfrac{\lambda_{\text{max}}}{c} \, \left( \dfrac{r_{_{\text{L}}}}{\lambda_{\text{max}}} \right)^{2-q} 
\end{equation}

It scales with the particle momentum--energy as $t_{_{\text{FII}}} \propto p^{2-q}$. The quasi-linear approach of \cite{schlickeiser1989} provides rather accurate (with an order of magnitude precision compared to numerical test-particle simulations) description of stochastic particle acceleration in the case of non-relativistic Alfv\'en speeds $\beta_\text{A} \ll 1$ and low turbulence levels $\delta B \ll B_0$, and for mildly relativistic Alfv\'en waves and turbulence levels comparable to unity \citep{osullivan}.    

In the case of `hard-sphere' turbulence, the stochastic acceleration time-scale $t_{_{\text{FII}}}$ is energy-independent, with

\begin{equation*} 
    D_{_{p,\text{FII}}}(p) \, \approx \, \beta_\text{A}^2 \, \left(\frac{\delta B}{B_0}\right)^2 \, \left(\frac{\lambda_{\text{max}}}{c}\right)^{-1}  p^2 \, \, \equiv \, \, D_0 \, p^2  
\end{equation*}

and

\begin{equation} \label{eq:stochacctscaleturblevel}
    t_{_{\text{FII}}} \, = \, \frac{1}{\beta_\text{A}^2} \, \left(\frac{B_0}{\delta B}\right)^2  \, \frac{\lambda_{\text{max}}}{c} \, = \, \frac{1}{D_0}    
\end{equation}

As one can see, the Fermi-II time-scale is controlled by the turbulence level. In case it varies with time due to the evolution of the turbulence, the Fermi-II time-scale also changes with time. To estimate $\beta_\text{A}$, the energy density can be evaluated directly from the electron spectrum, assuming that the relativistic electrons dominate the total energy density $\varepsilon$ and that for ultra-relativistic particles the pressure $P = \frac{1}{3} \varepsilon$. With these assumptions, the Alfv\'en speed for the case of relativistic MHD is given by \citep{gedalin}

\begin{equation} \label{eq:alfvenspeed} 
    \beta_\text{A} = \dfrac{1}{\sqrt{1 + \frac{4 \mu_0 \varepsilon}{3 B_0^2}}}
\end{equation}

The process of stochastic acceleration of electrons in the case of `hard-sphere' turbulence is described in the kinetic equation (Eq.~\ref{eq:kineticeqgeneral}) by two terms, the first one $-\dfrac{\partial}{\partial \gamma}\left(2 D_0 \gamma N_\text{e}(\gamma,t)\right)$ is due to drift of electrons to higher Lorentz factors, with $2 D_0 \gamma$ being proportional to particle energy gain per unit of time, and the second one $\dfrac{\partial}{\partial \gamma}\left(D_0\gamma^2 \dfrac{\partial N_\text{e}(\gamma,t)}{\partial \gamma}\right)$ describes the diffusion of the electron distribution in Lorentz factor space, where $D_0 \, \gamma^2$ is the energy diffusion coefficient. 

Just like the Fermi-I process, the Fermi-II process in our model can be activated and deactivated when necessary. When the acceleration by turbulence is not active, the $D_0$-term is set to zero. This introduces again an additional free parameter $t_{_{\text{dur,FII}}}$, which is the duration of the turbulent acceleration phase. In sub-section~\ref{subsec:twozonemodel} this parameter is further decomposed into two parameters $t_{\text{turb,r}}$ and $t_{\text{turb,d}}$, the rise and decay times of the turbulence.       

Fermi-II acceleration mechanism is a universal process expected to operate in relativistic jets. In particular, it could work inside or in the vicinity of the emitting zone. Spontaneous formation of turbulence in the magnetised plasma of the blob or surrounding it results in turbulent magnetic fields. An attractive feature of the Fermi-II process, is that it is capable of producing electron spectra $dN_\text{e} / d\gamma \propto \gamma^{-\alpha_{\text{es}}}$ with an index much harder than $\alpha_{\text{es}} = 2$ \citep{virtanen2005}. Also, the turbulent acceleration can serve as an efficient mechanism for re-acceleration of pre-accelerated particles.

\subsection{Escape and injection of particles}

The term $\dfrac{N_\text{e}(\gamma,t)}{t_{\text{esc}}}$ in the Eq.~\ref{eq:kineticeqgeneral} describes the escape of particles from the blob, with a characteristic escape time-scale $t_{\text{esc}}$, which in general case depends on the particle energy. In the case of negligible or weak turbulence in the emission zone, particles escape it freely, and the escape time-scale is $t_{\text{esc}} \sim 1 \, R_\text{b}/c$. Particles that undergo stochastic acceleration escape the region at a longer time-scale, due to diffusion process. The spatial diffusion coefficient and the momentum diffusion coefficient are linked as $D_x \, D_p \, \approx \, \beta_\text{A}^{2} \, p^2$ \citep{skilling}. The escape time-scale from a turbulent region is therefore related to the stochastic acceleration time-scale as \citep{tramacere}

\begin{equation} \label{eq:tescrelattotacc}
    t_{\text{esc}}^{(\text{turb})} = \frac{R_\text{t}^2}{c^2 \beta_\text{A}^2 \, t_{_{\text{FII}}}}
\end{equation}

where $R_\text{t}$ is the size of the turbulent zone. For `hard-sphere' turbulence, $t_{\text{esc}}$ becomes energy-independent:

\begin{equation}
\label{eq:tescedmf}
    t_{\text{esc}}^{(\text{turb})} = \left(\frac{R_\text{t}}{c}\right)^2 \, \left(\frac{\delta B}{B_0}\right)^2 \, \frac{c}{\lambda_{\text{max}}} 
\end{equation}

Finally the term $Q_{\text{inj}}(\gamma,t)$ is the number of particles injected in a unit volume per unit time and per unit of Lorentz factor interval. It can be implemented as time-dependent injection spectrum. Particles may be injected continuously over a given time interval, or in an impulsive manner.

\subsection{Numerical implementation}
\label{subsec:numericalimplement}

Following the kinetic approach discussed above, we developed a numerical code `EMBLEM' (Evolutionary Modelling of BLob EMission) for time-dependent modelling of blazar emission during flares. We use a fully implicit difference scheme by \cite{changcooper} to numerically solve the kinetic equation (Eq.~\ref{eq:kineticeqgeneral}) and retrieve the time evolution of the electron spectrum on a time-grid. For the case without any Fermi-II acceleration term, we use a particular case of the Chang and Cooper scheme described in \cite{chiaberge1999}. We impose boundary conditions ensuring particle conservation in case of no injection or escape, following \cite{parkpetrnum}. These boundary conditions prohibit leak of particles through the lower and upper boundaries of the Lorentz factor space that might occur due to processes merely changing particle energy (acceleration and cooling). As a result, the number of particles in the system is determined only by the injection and escape processes, while the acceleration and cooling processes cause particles to migrate between energy bins and cannot push particles beyond the user-defined domain of Lorentz factors. Throughout the paper, we define this domain wide enough in order to not artificially restrict the particle energy gain or loss processes. The SED of the emission from the blob is calculated at each time step. For this, the synchrotron emissivity and synchrotron self-absorption coefficient are evaluated using the expression from \cite{chiaberge1999}; the synchrotron intensity as a solution of the radiative transfer equation for the case of spherical geometry is adopted from \cite{katarzynskistat}. The inverse Compton emission is computed following the approach by \cite{katarzynskistat}, which includes the full Klein-Nishina cross-section \citep{jones1968}. Finally, we transform the emission from the blob reference frame to the observer's frame. We adopt a value for the Hubble constant of $H_0 = 70$ km s$^{-1}$ Mpc$^{-1}$. The absorption of $\gamma$-rays due to their interaction with the Extragalactic Background Light (EBL) is taken into account, with the use of a publicly available module \footnote{\url{https://github.com/me-manu/ebltable}}. For the current application, we use the EBL model by \cite{dominguez}. The light curves are calculated by integrating over the time dependent emission in the energy range of interest.

The treatment of the evolution of the electron spectrum was verified using test equations from \cite{parkpetrnum}, as well as by comparing to analytical solutions for simple cases of the kinetic equation. The description of the radiative emission was cross-checked with the output of the code by \cite{cerruti2015}.

\section{A model for the low-state emission of Mrk\,421}
\label{sec:scenario_ls}

In this paper, we consider that flaring activity is not simply superposed to the steady emission of the source, but caused by a perturbation of its quiescent (steady) low-state. This assumption results in significant additional constraints on the VHE emitting zone. Various effects can induce such a disturbance, including variable injection rate (e.g.\ \cite{mk1997}), a passing shock (e.g.\ \cite{marschergear1985} ; \cite{sikora2001}), various instabilities (e.g.\ \cite{meliani} ; \cite{kink}), even stars crossing the jet \citep{barkov2012}, etc. We will focus on scenarios where particle acceleration processes (Fermi-I and Fermi-II) are responsible for launching the flares. For this purpose, we first model the steady state of Mrk\,421, and then incorporate perturbations and simulate the flare. 

To describe the low state, we use the data set from \cite{abdo2011}, in which authors present the SED of Mrk\,421 averaged over the observations taken during the MWL campaign from January 19 to June 1, 2009, when the source showed a very low level of activity. We consider that the low-state measurement by \cite{abdo2011} represents the SED of the quiescent emission of the source on a relatively long-term time range. As the low-state observations and February 2010 flare are separated by only about 1 year, we assume that the VHE blob did not undergo significant adiabatic expansion during this period and so that the broad-band emission of the source in the pre- and post-flare state in February 2010 is described by the low-state measurement by \cite{abdo2011}. The validity of this assumption is confirmed by the close match between the simulated quiescent source fluxes and observed pre- and post-flare fluxes in different energy bands.

\begin{figure}
\centering
\includegraphics[height=75mm]{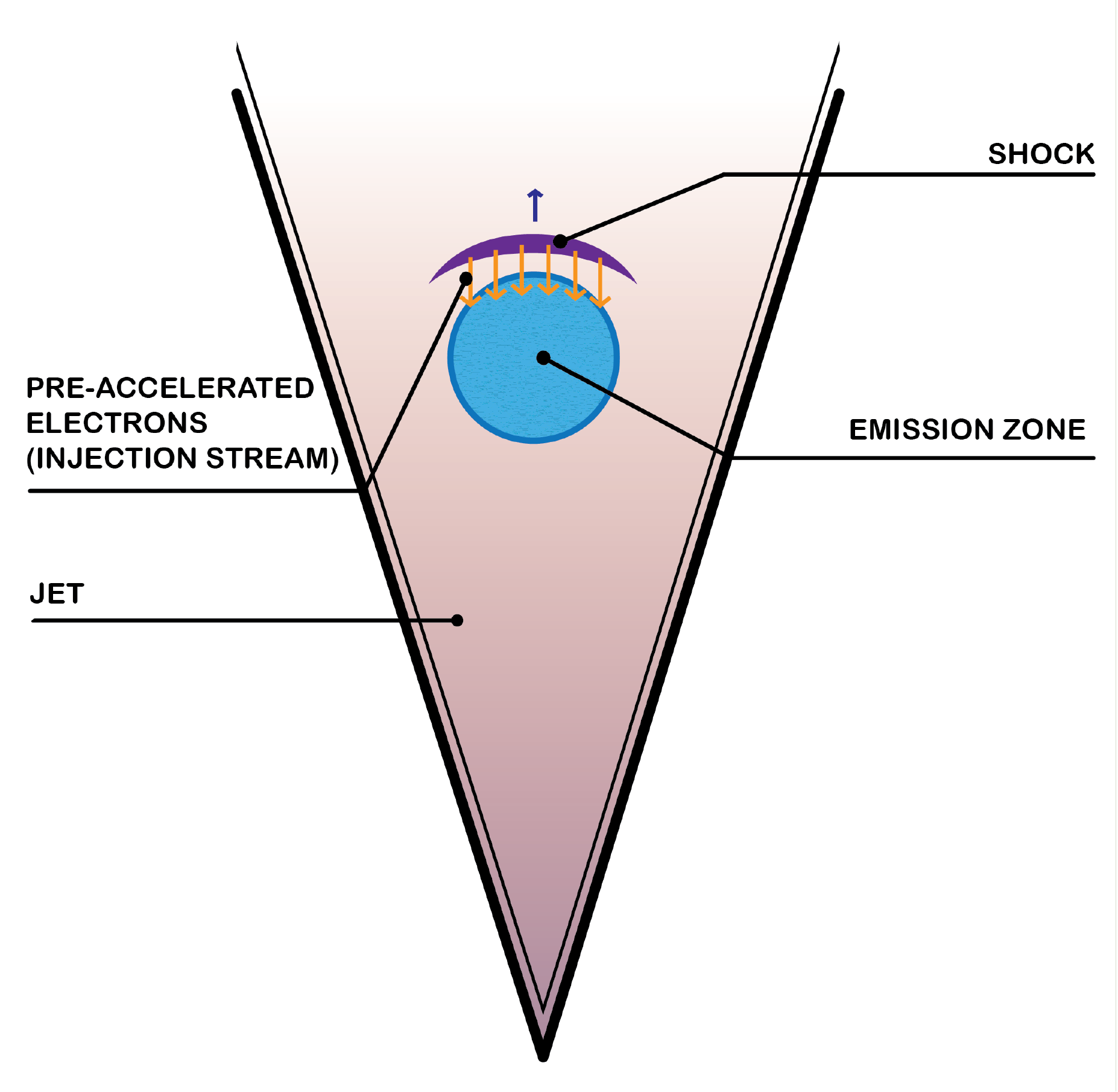}
\caption{Scheme illustrating a physical scenario for the long-term low-state emission of Mrk\,421. The blue filled circle represents the VHE $\gamma$-ray emitting zone (the blob) moving along the jet. The violet curve indicates the stationary shock leading the blob, which continuously accelerates particles of the upstream plasma and injects them into the downstream emission region. This injection flux is displayed with the orange arrows. The continuous influx of electrons pre-accelerated by the Fermi-I mechanism is responsible for the long-term quiescent emission of the source. Particles injected into the emitting zone radiate and cool in accordance with the SSC scenario, and escape from the blob at a time-scale $t_{\text{esc}} = 1 \, R_\text{b}/c$.}
\label{fig:steadystatesketch}
\end{figure}

We suppose that pre-accelerated electrons are continuously injected into the emission zone in the form of a steady `stream' (see Fig.~\ref{fig:steadystatesketch}). Taking into account the analysis and conclusion by \cite{yan2013} on the low-state of Mrk\,421, we assume that the spectrum of the injected electrons $Q_{\text{inj,0}}(\gamma)$ is a power law with an exponential cutoff, resulting from a shock acceleration process. We also suppose that the electrons are injected only above a certain Lorentz factor $\gamma_{\text{min,inj}}$.

\begin{figure*}
\centering
\includegraphics[width=155mm]{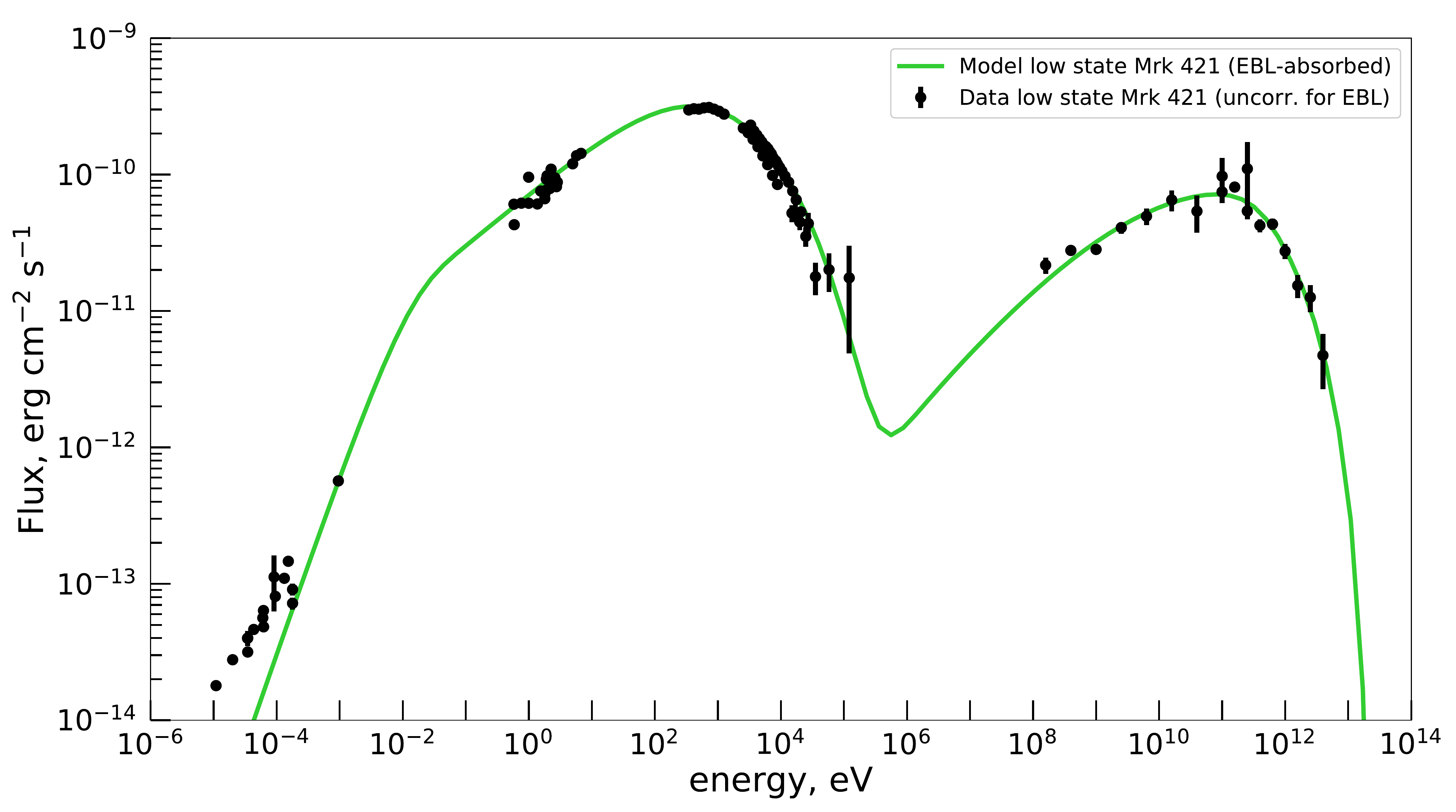}
\caption{Modelling of the spectral measurement of the long-term low-state emission of Mrk\,421. The green curve represents the model -- SED in the asymptotically established steady state with EBL absorption included \protect\citep{dominguez}, and black points show the MWL data set from \protect\cite{abdo2011}. The observed $\nu  F_{\nu}$ flux in the VHE range is uncorrected for EBL absorption. The host galaxy contribution has been subtracted from the optical measurements, the data in the optical-to-X-ray range were corrected for the effect of Galactic extinction.}  
\label{fig:qsm}
\end{figure*}

We simulate the steady state of the source with our EMBLEM code assuming an escape time-scale of $t_{\text{esc}} = 1 \, R_\text{b}/c$ from the blob. The population of electrons in the emitting region radiates and cools in accordance with the SSC scenario. The low state of Mrk\,421 in our description is the asymptotically established equilibrium between the gain processes (injection) and losses (cooling and escape), and corresponds to the stationary solution of the kinetic equation (Eq.~\ref{eq:kineticeqgeneral}), $\frac{\partial N_\text{e}}{\partial t} = 0$. The blob radius $R_\text{b}$ and Doppler factor $\delta_\text{b}$ are constrained in a way that the related variability time-scale of the source derived from the causality arguments, $t_{\text{var}} \sim R_\text{b} (1+z) \, (c \delta_\text{b})^{-1}$, is of the order of 1 d. This value is chosen since the February 2010 flare, considered to be a perturbation of the steady state, proceeds at a similar time-scale. Also, the variability time-scale of 1 d is used by \cite{abdo2011} in their instantaneous modelling of the low-state data set. We vary physical parameters of the source until we satisfactorily fit the MWL data set (cf. Table~\ref{tab:lowstateparams}). The parameters are consistent with the results of the instantaneous modelling presented by \cite{abdo2011}. The characteristic variability time-scale in our model is $t_{\text{var}} \approx 0.4$ d. Fig.~\ref{fig:qsm} displays the MWL measurements from \cite{abdo2011} together with our steady-state model, which shows a very good agreement with the data. The observed radio emission in the energy range $10^{-5} \leq E_{\gamma} \leq 10^{-3}$\,eV is thought to be dominated by the synchrotron emission of the extended jet, which is not included in our radiative model.

\begin{table*}
\footnotesize
\centering
\begin{tabular}{|c|c|c|c|} 
 \hline
 Parameters of quiescent state & Symbol & Our model & \cite{abdo2011} \\ 
 \hline
  Magnetic Field [G] & $B$ & $0.04$ & $0.038$  \\ 
 \hline
  Comoving blob radius [cm] & $R_\text{b}$ & $2.8 \times 10^{16}$ & $5.2 \times 10^{16}$ \\
 \hline
 Doppler Factor & $\delta_\text{b}$ & $29$ & $21$  \\ 
 \hline
 Time-scale of electron escape & $t_{\text{esc}}$ & $1 \, R_\text{b}/c$ & not defined \\
 \hline
 Spectrum of injected electrons & $Q_{\text{inj,0}}(\gamma)$ & \begin{tabular}{@{}c@{}} 
 $A_{\text{inj}} \, \gamma^{-\alpha_{\text{inj}}} \, \text{exp}(-\gamma/\gamma_{\text{cut}})$ 
 , \, for $\gamma \geq \gamma_{\text{min,inj}}$ \\ 
 0 , \, for $\gamma < \gamma_{\text{min,inj}}$ \end{tabular}
  & not defined \\
 Injection spectrum normalisation [cm$^{-3}$ s$^{-1}$] & $A_{\text{inj}}$ & $2.63 \times 10^{-3}$ & \\
 Injection spectrum slope & $\alpha_{\text{inj}}$ & $2.23$ & \\
 Min. Lorentz Factor in inj. spectrum & $\gamma_{\text{min,inj}}$ & $800$ & \\
 Cutoff Lorentz Factor in inj. spectrum & $\gamma_{\text{cut}}$ & $5.8 \times 10^5$ & \\
 \hline
\end{tabular}
\caption{Physical parameters of the Mrk\,421 low state. 3$^{\text{rd}}$ column: our scenario, 4$^{\text{th}}$ column: instantaneous modelling by \protect\cite{abdo2011}}
\label{tab:lowstateparams}
\end{table*}

It should be noted that the assumed injected electrons can be accelerated by a relativistic electron-positron shock as the inferred slope of their spectrum, $\alpha_{\text{inj}} = 2.23$, is very close to the theoretical predictions for such type of shocks (e.g.\ \cite{sironireview2015}). The most natural solution is therefore to consider that the injection flow is due to a stationary relativistic shock in front of the blob, generated by the interaction of the upstream extended jet plasma with the blob (see Fig.~\ref{fig:steadystatesketch}). In such a type of scenario \citep{kirkmastichiadis}, the particles of the upstream plasma would be passing through the shock, undergoing Fermi-I acceleration, and being injected downstream into the emitting blob. The cutoff at high energies in the injection spectrum may result either from the limited power of the shock accelerator (Hillas criterion for the maximal attained energy of a particle), or from a drop in efficiency of shock acceleration of particles with Lorentz factor above $\gamma_{\text{cut}}$ when their Larmor radius becomes larger than the characteristic size of turbulent eddies in the medium downstream of the shock. The injection of electrons only above $\gamma_{\text{min,inj}} = 800$ can be explained by their pre-acceleration to these Lorentz factors prior to the shock, e.g.\ in the vicinity of the central engine.

\section{A general validity criterion for one-zone flaring scenarios with a transient shock}
\label{sec:generalvalcrit}

\begin{figure*}
\centering
\includegraphics[height=75mm]{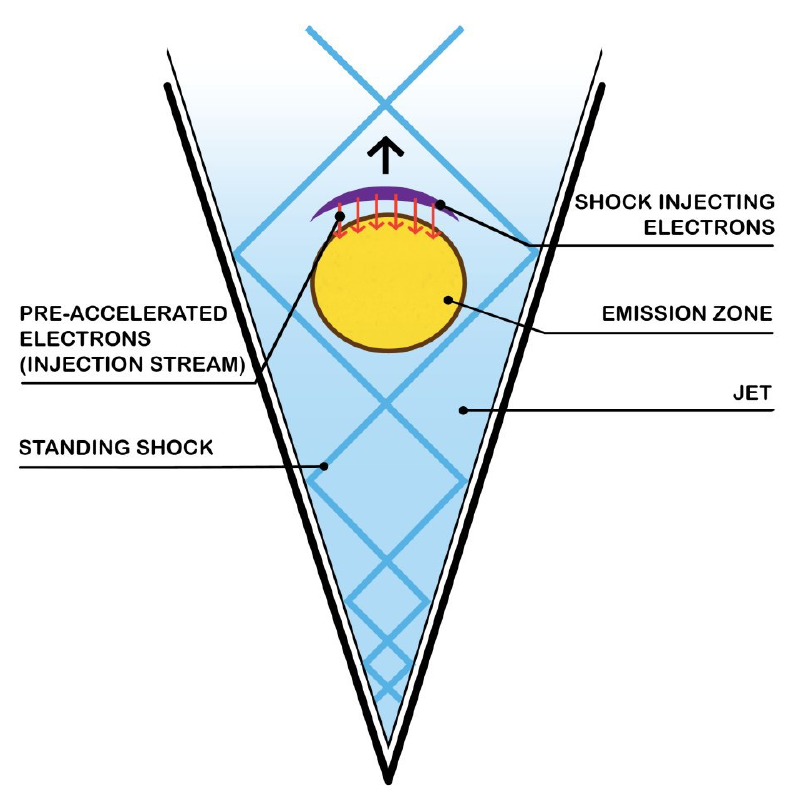} \hspace*{15mm}
\includegraphics[height=75mm]{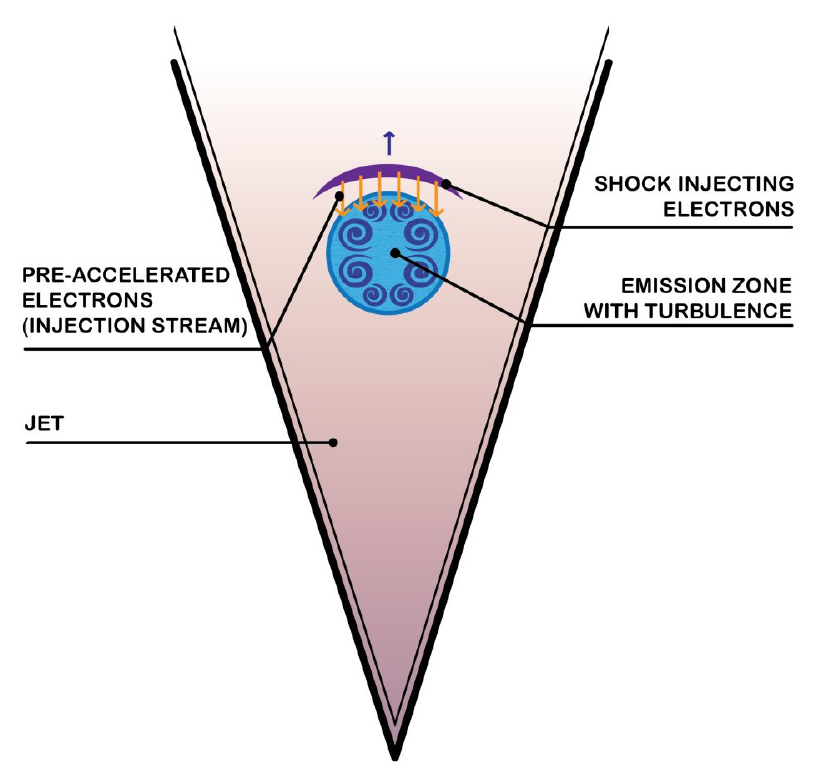}
\caption{Sketches illustrating possible one-zone scenario for flares. The violet curve indicates the steady shock in front of the blob, which accelerates upstream particles and supplies the electron stream injected in the blob for the stationary low-state emission (shown in orange arrows). Left: sketch in which the observed flux increase is due to a transient shock event perturbing the electron population in the emitting region. For instance, a standing shock with a so-called `diamond structure' is present in the jet. The blob is passing through the knot of such a standing shock and the electron population is (re-)accelerated by transient Fermi-I process. Right: sketch in which the quiescent blob is perturbed by turbulence induced for instance by extended jet inhomogeneities and the flare is triggered by stochastic acceleration of particles.}
\label{fig:sketchonezone}
\end{figure*}

To describe the flare emission, a one-zone model is the most basic scenario that can be tested. A single emission region is then responsible for both the low-state emission and the short-term flaring emission, caused by a perturbation in the emitting blob. Strong flux rise in the VHE part of the spectrum can be the signature of enhanced particle acceleration while the emitting blob is disturbed by e.g.\ a passing shock and turbulence which induces a `boost' or hardening of the steady-state particle spectrum. In the rest of this section we focus in detail on the scenario in which the flare is induced by a shock passing through the emitting blob (Fig.~\ref{fig:sketchonezone}, left panel).

Following our main assumption, physical parameters of the emitting zone do not significantly change on average during the passage of the shock, namely magnetic field, size, Doppler factor, escape time, as well as injection function, and we neglect the inverse Compton cooling so that the cooling rate is constant in time. The particles inside the blob are re-accelerated by the transient shock, the steady-state electron spectrum $N_{\text{e},0}(\gamma)$ is then perturbed and the blob emission as well. We show hereafter that when this scenario reproduces a flux increase observed at some frequency, for instance in the X-rays, the resulting flux induced at another frequency, for instance in the optical, can be or not perturbed just at the right level to fit the data, which can confirm or reject the model. We first need to connect the evolution of the electron spectrum to the flux variations for a generic case, then determine the general form of a time-dependent electron spectrum disturbed by a transient shock. Then using the derived general form of the electron spectrum we fit the X-ray data and find the shock acceleration time-scale, and as a final step we predict the optical flux increase to be compared to the observed one.            

To explore how the photon flux enhancement is linked to the electron spectrum time evolution, we consider here only synchrotron emission variations (e.g.\ for Mrk\,421 from radio band to hard X-rays), using the $\delta$-approximation which considers that an electron with a Lorentz factor $\gamma$ emits only at its critical frequency corresponding to the photon energy (in the observer's frame) $E_{\gamma} \simeq 5 \times 10^{-9} \, B_{\text{G}} \gamma^2 \delta_{\text{b}} \, (1+z)^{-1}$ \citep{rybickylightman}, where $B_{\text{G}}$ is the magnetic field in Gauss. The synchrotron SED is then given by (e.g.\ \cite{dermerschlickeiser2002}) 

\begin{equation} \label{eq:synseddeltaapp}
     E^2 \, \dfrac{dN_{\text{ph,syn}}}{dE}(E) \, \propto \, \bar{\gamma}^3 \, N_\text{e}(\bar{\gamma})
\end{equation}

where $\bar{\gamma} = \sqrt{E/\kappa}$, which is the Lorentz factor of an electron emitting synchrotron photon with energy $E$, and $\kappa = 5 \times 10^{-9} \, B_{\text{G}} \delta_{\text{b}} \, (1+z)^{-1}$.

The energy and the photon flux of a light curve in the energy range from $E_{\text{min}}$ to $E_{\text{max}}$ are given by

\begin{equation*} 
    F_{\text{erg}} \propto \int_{E_{\text{min}}}^{E_{\text{max}}} E \, \dfrac{dN_{\text{ph,syn}}}{dE}(E) \, dE \propto \int_{E_{\text{min}}}^{E_{\text{max}}} E^{1/2} \, N_\text{e}(\bar{\gamma}(E)) \, dE
\end{equation*}

\begin{equation*} 
    F_{\text{ph}} \propto \int_{E_{\text{min}}}^{E_{\text{max}}} \dfrac{dN_{\text{ph,syn}}}{dE}(E) \, dE \propto \int_{E_{\text{min}}}^{E_{\text{max}}} E^{-1/2} \, N_\text{e}(\bar{\gamma}(E)) \, dE
\end{equation*}

These expressions connect the electron spectrum variations to those of the energy and photon flux in a certain energy range. Using them, we write out the flux increase factor $\xi_{_{\text{LC}}}$, ratio of the peak flux in the light curve $F_{\text{peak}}$ to the quiescent flux $F_0$, 

\begin{equation} \label{eq:ratiofluxlcmacc}
    \xi_{_{\text{LC}}} \equiv \dfrac{F_{\text{peak}}}{F_0} = \dfrac{ \int_{E_{\text{min}}}^{E_{\text{max}}} E^s \, N_{\text{e,peak}}(\bar{\gamma}(E)) \, dE }{ \int_{E_{\text{min}}}^{E_{\text{max}}} E^s \, N_{\text{e},0}(\bar{\gamma}(E)) \, dE }
\end{equation}

where $s = 1/2$ for the energy flux ratio, and $s = -1/2$ for the photon flux ratio.

For the spectral flux increase factor $\xi_{\text{spec}}$ at a particular photon energy $E$, a similar expression can be written using the Eq.~\ref{eq:synseddeltaapp}:

\begin{equation} \label{eq:ratiofluxmaxspec}
    \xi_{\text{spec}} \equiv \dfrac{dN_{\text{ph,syn,peak}}(E) / dE }{dN_{\text{ph,syn},0}(E) / dE } = \dfrac{N_{\text{e,peak}}(\bar{\gamma}(E))}{N_{\text{e},0}(\bar{\gamma}(E))} 
\end{equation}

To compute the flux increase factor, we now focus on the analytical derivation of the electron spectrum evolution during the passage of a shock through the emitting blob. $N_{\text{e},0}(\gamma)$ is the steady state electron spectrum and $N_{\text{e,FI}}(\gamma,t)$ the evolving one during the passage of the shock. The evolution of $N_{\text{e,FI}}(\gamma,t)$ is governed by the kinetic equation with the transient shock acceleration term, 

\begin{equation} \label{eq:kineqshockaccel}   
 		\dfrac{\partial N_{\text{e,FI}}(\gamma,t)}{\partial t} = \dfrac{\partial}{\partial \gamma}\left( W(\gamma) \, N_{\text{e,FI}}(\gamma,t)\right) \, - \, \dfrac{N_{\text{e,FI}}(\gamma,t)}{t_{\text{esc}}} \, + \, Q_{\text{inj}}(\gamma)
 		\end{equation}

with $W(\gamma) = b_\text{c} \gamma^2 - \gamma / t_{_{\text{FI}}}$.\\

This equation can be solved analytically as presented in the Appendix~\ref{appen:analyticalsolutionke}, so that  

\begin{multline} \label{eq:neshockpert}
    N_{\text{e,FI}}(\gamma,t) = N_{\text{e},0}(\gamma) \, + \, \int_{0}^{t} \dfrac{\Gamma(\gamma,t,t^{\prime}) \cdot \text{exp} \! \left[(1/t_{\text{esc}} \, - \, 1/t_{_{\text{FI}}}) \, (t^{\prime} \, - \, t)\right]}{b_\text{c} \, t_{_{\text{FI}}} \, \gamma^2} \, \cdot \\ \cdot \, \left[ Q_{\text{inj}}(\Gamma(\gamma,t,t^{\prime})) \, + \, \left(b_\text{c} \Gamma(\gamma,t,t^{\prime}) \, - \, 1 / t_{\text{esc}} \right) \, N_{\text{e},0}(\Gamma(\gamma,t,t^{\prime})) \right]  \, dt^{\prime}
\end{multline}

where

\begin{equation} \label{eq:steadystatesolut1}
    N_{\text{e},0}(\gamma) = \dfrac{1}{b_\text{c} \gamma^2} \int_{\gamma}^{\gamma_{\text{max}}} Q_{\text{inj}}(\gamma^{\prime}) \cdot \text{exp} \left( \frac{1/\gamma^{\prime} - 1/\gamma}{b_\text{c} t_{\text{esc}}} \right) d\gamma^{\prime}
\end{equation}

\vspace*{4mm}

and

\begin{equation} \label{eq:biggamma1}
    \Gamma(\gamma,t,t^{\prime}) = \dfrac{\gamma \cdot e^{(t^{\prime} - t)/t_{_{\text{FI}}}}}{1 \, + \, \gamma b_\text{c} \, t_{_{\text{FI}}} \, (e^{(t^{\prime} - t)/t_{_{\text{FI}}}} \, - \, 1)}
\end{equation}

The parameter $\gamma_{\text{max}}$ is the Eq.~\ref{eq:steadystatesolut1} represents the maximal Lorentz factor of the electron population in the blob. As already discussed in sub-section~\ref{subsec:numericalimplement}, throughout this paper, we set the Lorentz factor range in numerical computations wide enough (with large margins), allowing particles to freely migrate between the energy bins when they gain or lose energy. Therefore, we calculate the steady-state electron spectrum given by the Eq.~\ref{eq:steadystatesolut1} with $\gamma_{\text{max}} \rightarrow \infty$.

Knowing how the electron spectrum evolves during the passage of the shock, we can now evaluate the expected flux increase at different wavelengths. First, we compute the steady-state electron spectrum from Eq.~\ref{eq:steadystatesolut1}, using the fit of the steady-state SED of the source which one should obtain using the approach described in Section~\ref{sec:scenario_ls} with the physical parameters (including B, $\delta_\text{b}$, $t_{\text{esc}}$) and the injection function $Q_{\text{inj}}(\gamma)$. The duration during which the transient shock acceleration is active is given by the crossing time of the shock, which can be approximated by the observed rise time of the flare, corrected for relativistic effects, i.e.\ $t_{\text{cs}} \approx t_{_{\text{rise}}} \, \delta_\text{b}$. We then find the electron spectrum at the flare peak (at the moment when the shock exits the blob) depending on the Fermi-I time-scale by evaluating $N_{\text{e,FI}}(\gamma,t)$ (Eq.~\ref{eq:neshockpert}) at the moment $t = t_{\text{cs}}$: $N_{\text{e,peak}}(\gamma ; t_{_{\text{FI}}}) = N_{\text{e,FI}}(\gamma,t=t_{\text{cs}} ; t_{_{\text{FI}}})$, the semicolon indicates separation between the arguments (Lorentz factor, time) and the parameter (Fermi-I time-scale). The corresponding acceleration time-scale is then deduced from the observed synchrotron flux increase in the X-ray band using Eq.~\ref{eq:ratiofluxlcmacc}, in which we use the peak electron spectrum $N_{\text{e,peak}}(\gamma ; t_{_{\text{FI}}})$ obtained just above.

We solve numerically the equation for $t_{_{\text{FI}}}$, and obtain its value $t_{_{\text{FI}}}^{*}$ required to produce the observed enhancement of the flux in the X-ray light curve. In case X-ray spectral measurements are available for the flare peak, one could derive the Fermi-I time-scale using those data, solving numerically the Eq.~\ref{eq:ratiofluxmaxspec}, to avoid the need of integration over the photon energies.

Finally, with the retrieved acceleration time-scale and peak electron spectrum, one can deduce from Eq.~\ref{eq:ratiofluxlcmacc} or Eq.~\ref{eq:ratiofluxmaxspec} the expected flux increase in the optical band and compare it to the one reconstructed from the optical data (after subtraction of the host galaxy contribution). In case of a significant divergence between the two values, one needs to conclude that the one-zone model with a shock traversing the emitting blob cannot satisfactorily describe the MWL data set, and reject the scenario.

Indeed, it is sufficient to retrieve only one parameter, $t_{_{\text{FI}}}$, to apply the criterion when one knows the physical parameters of the low state and the duration of the flux rise, and when the optical and X-ray fluxes at the very peak are available. Conversely when there is no precise information on the time of the flux increase, there are two unknown parameters and an additional relation is needed. This can be achieved for instance if the flux increase ratio is available in both the soft and the hard X-ray bands, since one can then deduce $t_{\text{cs}}$ and $t_{_{\text{FI}}}$ by solving numerically a system of two equations, namely applying Eq.~\ref{eq:ratiofluxlcmacc} with $N_{\text{e,peak}}(\gamma) = N_{\text{e,FI}}(\gamma ; t=t_{\text{cs}} , t_{_{\text{FI}}})$ to the soft and to the hard X-ray fluxes. Then one can predict the factor of the optical flux increase by applying the two inferred parameters $t_{\text{cs}}$ and $t_{_{\text{FI}}}$.

\section{Scenarios for the flare emission of Mrk\,421}
\label{sec:scenario_flare}
In this section, we explore whether the February 2010 flare of Mrk\,421 can be interpreted as being due to a moderate and non-destructive perturbation of the quiescent VHE emission region of the source. One-zone scenarios which provide the simplest way to connect low-state emission to the flaring one are considered first. There are then only two main free parameters to describe the flaring state, the time-scale of the shock or stochastic acceleration processes $t_{_{\text{FI/II}}}$, and the duration of the acceleration phase $t_{_{\text{dur,FI/II}}}$ related to the rise time of the flare, which strongly constrains the picture. Two-zone scenarios are then developed to better account for the complexity of the observed MWL time evolution.

\subsection{One-zone models}
\label{subsec:onezonemodels}

We first attempt to fit the flare data set by perturbing the low-state with a crossing shock or with turbulence, strictly within the central blob radiating the steady VHE emission. According to the assumption of weak perturbation, macroscopic physical parameters describing the steady state are kept constant at first order.   

\subsubsection{Analytical results: a passing shock}

In this case, the flare is simply initiated by a shock crossing the emitting blob, perturbing the particle population. The general criterion obtained in  Section~\ref{sec:generalvalcrit} can be directly applied considering the X-ray light curve in the energy range between 0.5 and 2 keV, and the optical {\it V}-band `light curve' (host galaxy subtracted). The flux increase ratio for the X-ray light curve  $\xi_\text{x} = F_{\text{x,peak}} / F_{\text{x},0} \approx 3.7$ and the rise time of the X-ray flare is around 3 to 4 d (in the observer's frame). We therefore assume an average value $t_{\text{cs}} \approx 101.5$ d, which translates into a shock speed relative to the blob of the order of  $\beta_{\text{sh}} \approx 0.1$, and recover $t_{_{\text{FI}}} \approx 1.65 \, R_\text{b} / c \approx 17.8$ d. from the Eq.~\ref{eq:ratiofluxlcmacc} for the observed value of $\xi_\text{x}$. Indeed the final result on $t_{_{\text{FI}}}$ depends quite weakly on the duration of the shock passage. 

\begin{figure*}
\centering
\includegraphics[height=50mm]{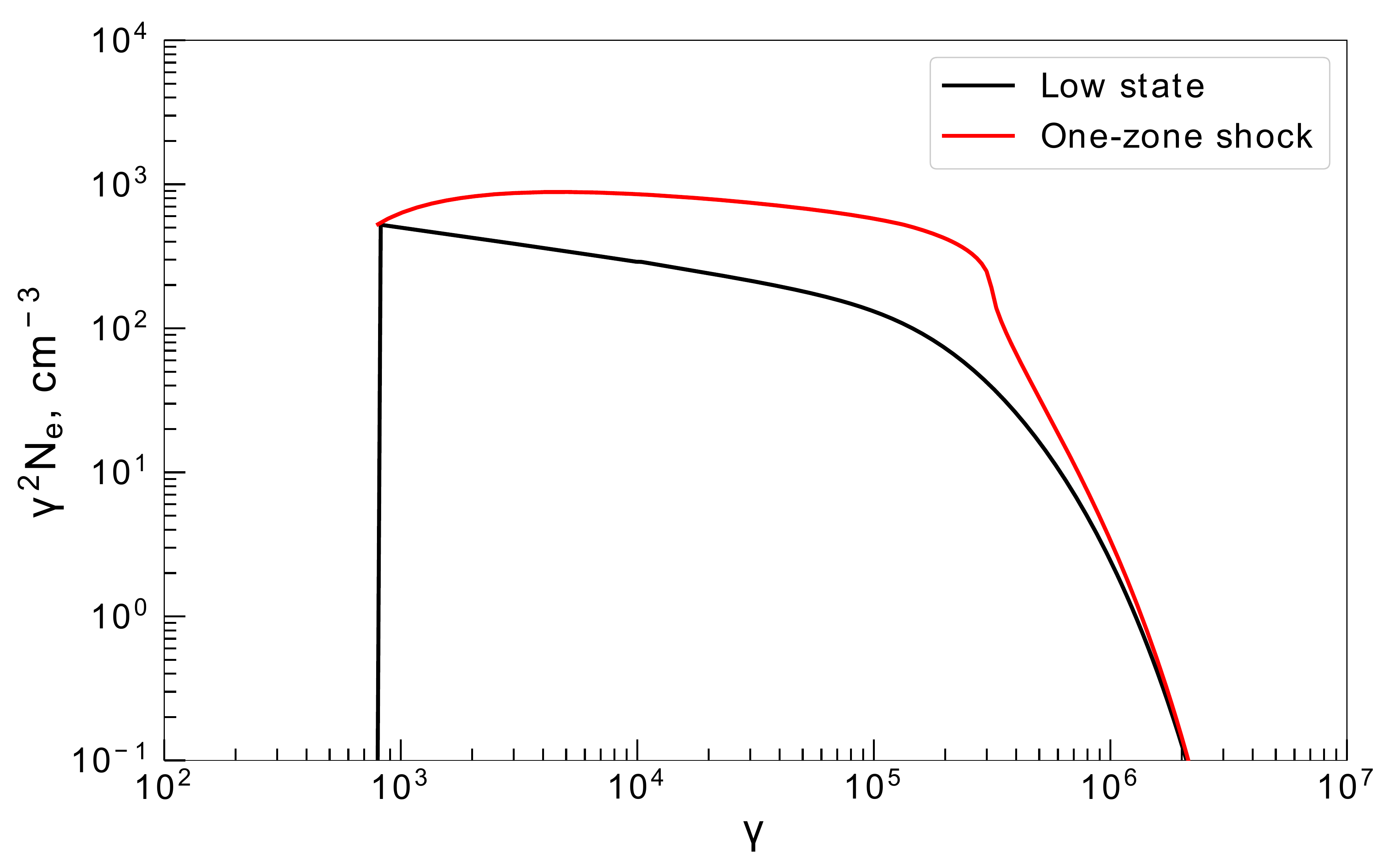} \hspace*{1mm} \includegraphics[height=50mm]{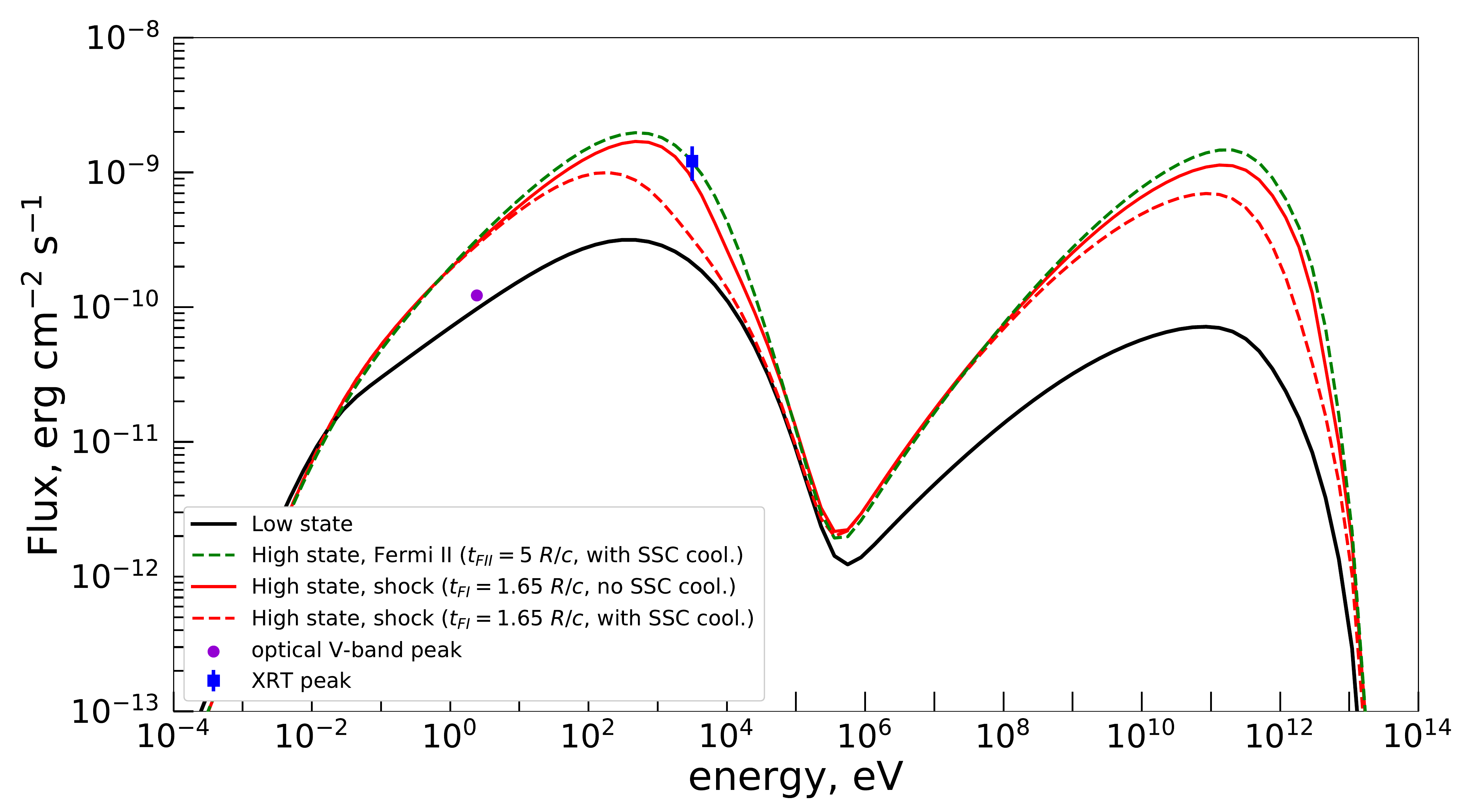}
\caption{Simulated electron spectrum and SEDs for one-zone flare scenario. Left: analytical calculation (using Eq.~\ref{eq:neshockpert}) of the electron spectrum disturbed by a shock with $t_{_{\text{FI}}} = 1.65 \, R_\text{b} / c$ at the moment of the peak of the flare (red), compared to the steady-state electron spectrum (black). The inverse Compton cooling effect is not included for both spectra. Right: peak SEDs for shock (dashed red curve) and Fermi-II (green curve) acceleration processes perturbing the blob simulated with the numerical code, with superimposed optical and X-ray spectral measurements near the flare peak. Black curve shows the steady-state SED. Solid red curve indicates the SED corresponding to the analytical electron spectrum shown in the left panel (full SED computation, inverse Compton cooling neglected), dashed red curve represents the same as the solid one but with the inverse Compton cooling effect taken into account. Green curve displays the SED for the model in which Fermi-II process with $t_{_{\text{FII}}} = 5 \, R_\text{b}/c$ is acting on the blob population (inverse Compton cooling is taken into account). The Fermi-II time-scale is tuned in a way that the peak SED fits the X-ray data. For all the models, the acceleration process acts for 3.5\,d in the observer's frame ($t_{_{\text{dur,FI/II}}} = 101.5$ d in the frame of the blob). One can see that all one-zone models with shock or stochastic acceleration fitting the X-ray flare overproduce emission at the optical wavelengths.} 
\label{fig:onezoneshockturb}
\end{figure*}

Eq.~\ref{eq:ratiofluxlcmacc} now applied to the optical light curve shows that such passing shock induces a flux increase ratio $\xi_{\text{opt}} \approx 3.3$ in the optical band, which is much higher that the observed value $\xi_{\text{opt,obs}} \approx 1.26$. Thus, the shock needed to reproduce the X-ray flare perturbs too much the optical flux, and one can conclude that this scenario is not satisfactory. This is illustrated in Fig.~\ref{fig:onezoneshockturb}. The left panel shows the electron spectrum, disturbed by the shock with Fermi-I acceleration and passage time-scales deduced from the X-rays. We verified that the analytical calculation of the electron spectrum (using Eq.~\ref{eq:neshockpert}) appears in close agreement with the  numerical results obtained with the EMBLEM code. The right panel of Fig.~\ref{fig:onezoneshockturb} displays the SED associated to the perturbed electron spectrum. One can clearly see that the shock causes too high pile-up in the optical part of the spectrum. Even though the analytical approach disregards the inverse Compton cooling, it does not affect the conclusion: a stronger shock needed to compensate the inverse Compton cooling and still describe the X-ray data, would lead to even higher rise of the optical flux. Therefore the criterion presented in Section~\ref{sec:generalvalcrit} provides a lower limit on the optical flux increase.

\subsubsection{Numerical simulations results: shock and turbulence}

While the analytical approach already excludes the one-zone shock acceleration model, one-zone models where Fermi-II acceleration plays a role need to be tested with the full numerical code. We explore various acceleration time-scales for models with only turbulent acceleration or with a combination of shock and turbulent acceleration. The duration of the acceleration phase $t_{_{\text{dur,acc}}}$ is dictated by the rise time of the light curve in the source frame, which is fixed from the 3.5\,d as seen by the observer corrected by relativistic effects. In both cases, no combination of parameters succeed to produce a satisfactory fit of the data. For the case where only Fermi-II acceleration perturbs the blob, the one-zone model that provides a good representation of the X-ray flux increase while varying $t_{_{\text{FII}}}$  overshoots again the optical flux in the high state as can be seen in the right panel of Fig.~\ref{fig:onezoneshockturb} (green curve). Contrarily, we find that the model describing well the optical peak data undershoots significantly the X-ray measurements. If we further consider that the escape time of electrons can be longer than $1 \, R_\text{b}/c$ during the Fermi-II acceleration process (Eq.~\ref{eq:alfvenspeed} and \ref{eq:tescrelattotacc}), it becomes even more difficult to achieve a good fit: the optical flux is even more overproduced. The problem with the optical flux excess persists when combining the shock and turbulent particle acceleration: we did not find any set of the two acceleration time-scales $t_{_{\text{FI}}}$ and $t_{_{\text{FII}}}$ which mitigates the excess of the optical flux, and provides a reasonable description of the MWL peak data set. One-zone models with a moderate perturbation of the quiescent state appear too much constrained to explain the observed MWL flux variations during the flare.

\subsection{Two-zone model}
\label{subsec:twozonemodel}

As one-zone models fail to satisfactorily describe the data set, we consider that the low-state and flaring emission emanate from two connected regions. The overproduction of the optical emission present in one-zone models can then be avoided if the optical emission is dominated by the quiescent region, and the X-ray flare by a second transient region. Since the conditions necessary for emission up to very high energies need to be fulfilled in both zones, their basic physical parameters cannot be too different. We assume that the steady-state emission comes from a relatively large region inside the extended jet as described in Section~\ref{sec:scenario_ls} and refer to it as the `quiescent blob'. The flaring emission originates from a smaller region that we identify as the `flaring region', in contact with the quiescent blob. Both quiescent and flaring zones move relativistically along the jet with the same Doppler factor.

\cite{yan2013} find that an instantaneous one-zone SSC scenario with a log-parabola electron spectrum provides a better fit of the Mrk\,421 February 2010 peak data than with a power law electron spectrum. Their analysis suggests a turbulent re-acceleration process as the possible cause of the flare. Moreover, \cite{zheng}, invoking Fermi-II mechanism in their attempt to explain the February 2010 flare, conclude that the spectral and timing properties of the outburst are better described with the `hard-sphere' approximation than with other turbulence types. This provides some support to the assumption of `hard-sphere' turbulence ($q = 2$) adopted in sub-section~\ref{subsec:stochasticacc} for our modelling of the Fermi-II acceleration effect. Considering these results, we focus our attention on the Fermi-II acceleration, sustained by `hard-sphere' turbulence, as the process powering activity in the flaring region. Because of that, we will also refer to the flaring region as `turbulent region'. 

\subsubsection{Application to the February 2010 flare}

\begin{figure}
\centering
\includegraphics[width=80mm]{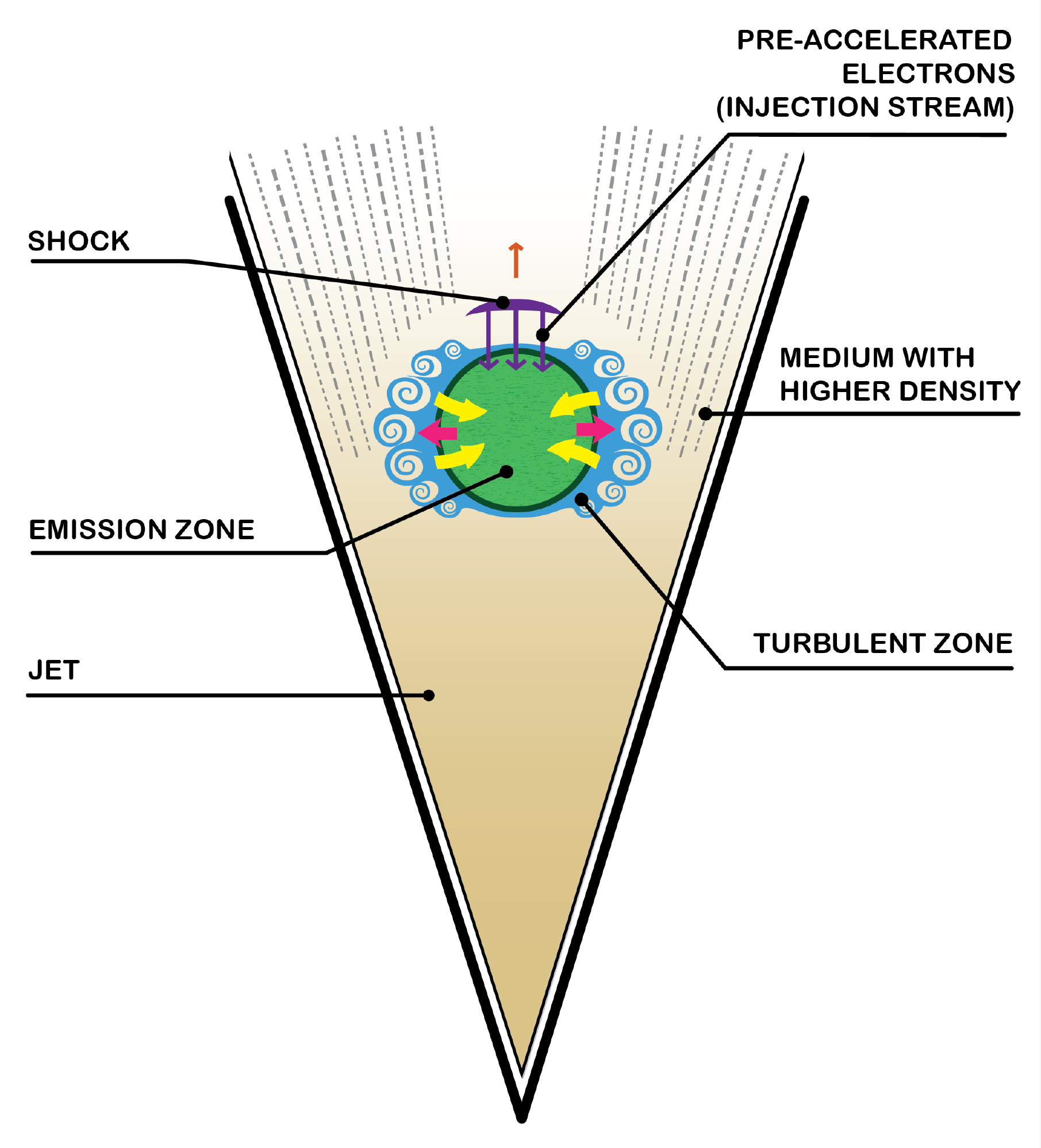}
\caption{Sketch showing a generic two-zone scenario for a flare in which a turbulent region is present around the quiescent emitting blob. The grey dashed lines show a material with higher density or different speed, which perturbs the medium around the blob and induces the turbulence. The violet curve above the quiescent blob represents the stationary shock wave, accelerating upstream electrons via Fermi-I process and injecting particles in the blob (injection flow indicated by violet arrows). The quiescent and turbulent regions exchange particles: ruby-coloured arrows display the injection of electrons escaping from the quiescent blob to the turbulent region, while yellow arrows represent the escaping particle flux from the turbulent region to the blob. The flow shown in yellow may be important or not (compared to the injection flow denoted by violet arrows), depending on the radii of the regions and on the respective escape time-scales. In the scenario we adopt in this paper, the particle flux from the turbulent region to the blob is negligible.}
\label{fig:sketchradturbulentzone}
\end{figure}

A generic configuration of the two-zone model we consider here is an abruptly appearing turbulent region at the interface of the quiescent blob and the surrounding jet. The two regions radiate and may exchange particles. There are two limiting cases: a steady-state emitting zone next to either (1) a non-radiative turbulent acceleration zone with an important particle escape, or (2) a radiative turbulent acceleration zone with negligible particle escape (see Fig.~\ref{fig:sketchradturbulentzone}). In the first case, the typical size of the turbulent zone has to be comparable to the one of the quiescent emission, and the magnetic field should be much lower. In the second case the magnetic field should be commensurate to the one in the blob and the typical size has to be smaller. We consider the radiative contribution of the surrounding extended jet to be negligible above the radio band, due for instance to significantly lower magnetic field and energetic particle density. After some trials, the second scenario proves to be the most promising one. Particles escaping from the quiescent blob are injected into the turbulent region, and are re-accelerated via the Fermi-II mechanism (see Fig.~\ref{fig:sketchradturbulentzone}). The electrons in the turbulent region radiate a flaring synchrotron and IC emission and the observed flux increases.

Following our fundamental assumption interpreting the flare just as a small perturbation of the quiescent state, we apply to the quiescent blob our modelling of the Mrk\,421 low state obtained in Section~\ref{sec:scenario_ls} with the physical parameters of Table~\ref{tab:lowstateparams}, and consider that the transient turbulent region should admit rather similar values. However there are now several new constraints and parameters for the physical description of the turbulent region. The effective size of the turbulent region, $R_{\text{tr}}$, is expected to be smaller than the one of the blob, $R_\text{b}$, so that the additional inflow of particles from the turbulent region is less important than the steady injection inflow. In order to have a significant emission from the turbulent region, its typical magnetic field $B_{\text{tr}}$ has to be sufficiently strong, i.e.\ comparable or higher to its strength inside the blob, and we first consider $B_{\text{tr}} \sim B \sim 0.04$ G as in Table~\ref{tab:lowstateparams}. The spectrum of particle injection in the turbulent region is defined by the spectrum of escaping particles from the quiescent blob, which is proportional to the steady-state spectrum of the electron population in the blob and assumed to be constant in time:

\begin{equation}
    \label{eq:injparticlesblobtoturb}
    Q_{\text{inj,qr-tr}}(\gamma) \simeq \dfrac{N_{\text{e},0}(\gamma)}{t_{\text{esc}}} \, \left(\dfrac{R_\text{b}}{R_{\text{tr}}}\right)^3 \, f_{\text{qr-tr}} 
\end{equation}

where $N_{\text{e},0}(\gamma)$ is the steady-state spectrum (Section~\ref{sec:scenario_ls}, and Eq.~\ref{eq:steadystatesolut1}), $t_{\text{esc}}$ is the escape time-scale from the quiescent emitting region (Table~\ref{tab:lowstateparams}), and $f_{\text{qr-tr}}$ is the fraction of particles escaping from the blob to the turbulent region.

The fraction $f_{\text{qr-tr}}$ depends on the detailed geometrical configuration of the two zones and of the stationary front shock. The turbulent region forms a kind of thick turbulent torus around the quiescent blob. For practical use we consider a simplified geometry and roughly describe the turbulent region as consisting of a few (namely 4 here) identical small spherical zones or `eddies' at the lateral edge of the quiescent blob. One eddy has a radius $R_{\text{ed}}$, which under our assumptions is related to the effective size of the turbulent region $R_{\text{tr}}$ via $R_{\text{tr}}^3 = 4 \, R_{\text{ed}}^3$. The contribution of each small eddy to $f_{\text{qr-tr}}$ is determined by the solid angle $\Omega$ enclosing it:

\begin{equation*}
    f_{\text{qr-tr}} = \dfrac{4 \cdot \Omega}{4\pi} \simeq \dfrac{1}{4\pi} \, 4 \, \dfrac{\pi R_{\text{ed}}^2}{R_\text{b}^2} = \left(\dfrac{R_{\text{ed}}}{R_\text{b}}\right)^2
\end{equation*}

Plugging this relation in Eq.~\ref{eq:injparticlesblobtoturb}, one obtains:

\begin{equation}
    \label{eq:injblobtoturb1sphere}
    Q_{\text{inj,qr-tr}}(\gamma) = \dfrac{N_{\text{e},0}(\gamma)}{4^{2/3} \, t_{\text{esc}}} \, \dfrac{R_\text{b}}{R_{\text{tr}}} = \dfrac{N_{\text{e},0}(\gamma)}{4 \, t_{\text{esc}}} \, \dfrac{R_\text{b}}{R_{\text{ed}}}
\end{equation}

The escape time-scale in the presence of turbulence is $t_{\text{esc,ed}} \propto \delta B^2 / (2 \mu_0)$ (see Eq.~\ref{eq:tescedmf}), and the Fermi-II acceleration time-scale is $t_{_{\text{FII}}} \propto \beta_\text{A}^{-2} \, (\delta B^2 / (2 \mu_0))^{-1}$ (see Eq.~\ref{eq:stochacctscaleturblevel}). The Alfv\'en speed is time-dependent, due to varying electron density in the turbulent region. A certain profile for the temporal evolution of the turbulence is needed to complete the description of the transient turbulent region, $\delta B^2 = \delta B^2(t)$.

The energy density contained in magnetic field fluctuations at a given time is determined by the balance between the injection of the turbulent energy in the flaring zone and its losses due to the work done on the acceleration of particles. Neglecting other losses such as damping of waves and the dependence of the losses on wavenumber, the equation governing the time evolution of the turbulent energy $U_{\text{turb}}(t) = \delta B^2(t) / (2 \mu_0)$ is \citep{burn}:

\begin{multline*}
    \dfrac{dU_{\text{turb}}(t)}{dt} \, = \, Q_{\text{turb}}(t) - \int_{\gamma_{\text{min}}}^{\gamma_{\text{max}}} \frac{2 \gamma m_\text{e} c^2}{t_{_{\text{FII}}}(t)} N_{\text{e,tr}}(\gamma,t) \, d\gamma \, = \\ = \, Q_{\text{turb}}(t) - \frac{2 \, \varepsilon(t)}{t_{_{\text{FII}}}(t)} 
\end{multline*}

Here $Q_{\text{turb}}(t)$ is the time-dependent rate of turbulent energy injection, $N_{\text{e,tr}}(\gamma,t)$ and $\varepsilon$ are the electron spectrum and the energy density of particles in the turbulent region. The integral from $\gamma_{\text{min}}$ to $\gamma_{\text{max}}$ corresponds to the systematic energy gain by particles due to the Fermi-II acceleration process, evaluated by integration of the corresponding term in the kinetic equation (Eq.~\ref{eq:kineticeqgeneral}) and represents the loss term for the turbulence. From Eq.~\ref{eq:stochacctscaleturblevel} and \ref{eq:alfvenspeed}, this loss term writes as

\begin{equation*}
\left(\dfrac{dU_{\text{turb}}(t)}{dt}\right)_{\text{loss}} \approx \dfrac{U_{\text{turb}}}{\lambda_{\text{max}} / c} 
\end{equation*}

The injection term $Q_{\text{turb}}(t)$ depends on the detailed physics of the turbulence generation. For practical use we express this complex term with a minimal number of free parameters. The simplest scenario we first consider is close to a gate function with a continuous injection of a turbulence that is constant in time over a given duration. In this case, the energy density of the turbulence grows from zero to a constant maximum level on a time-scale of $t_{\text{turb}} \approx \lambda_{\text{max}} / c$. Once the injection is stopped, particles extract all available energy from the reservoir on the same time-scale, with the energy density falling off exponentially. This time-scale corresponds to the decay time of the longest mode in the turbulent spectrum, and is the shortest possible time for the build-up or dissipation of the turbulence. However such temporal profiles of the turbulent energy induce flares with an extended plateau and do not generate the observed February 2010 flare shape. To achieve a better representation of the data, we suppose a non-constant profile of the injection with linear rise and decline on time-scales $t_{\text{turb,r}}$ and $t_{\text{turb,d}}$ respectively, both much longer than $\lambda_{\text{max}} / c$, so that the temporal behaviour of the turbulent energy density $\delta B^2(t) / (2 \mu_0)$ approximately replicates the behaviour of the injection function $Q_{\text{turb}}(t)$. This situation can correspond for instance to the central quiescent blob traversing a dense region with a density gradient. We further assume that the turbulent energy density is about equal to the energy density of the non-turbulent large-scale component of the magnetic field at the peak of the flare, $\delta B^2 |_{_{\text{peak}}} \, \sim \, B_{\text{tr}}^2$.

We can now proceed to compute the emission of the turbulent region. We model the emission of one eddy, and then scale its flux by a factor of 4 to estimate the total emission. After the turbulence has dissipated, the region dissolves in the ambient medium inside the jet, which we simulate by simply stopping particle injection. For a self-consistent description we evaluate the Alfv\'en speed (from  Eq.~\ref{eq:alfvenspeed}) and the escape and Fermi-II time-scales (from Eq.~\ref{eq:tescedmf} and Eq.~\ref{eq:stochacctscaleturblevel}) at each time step in the code. The escape of particles from the eddy at the beginning and at the end of the turbulence is considered as free-streaming with an escape time-scale $t_{\text{esc,ed},0} = 1 \, R_{\text{ed}} / c$, while during the turbulence phase the profile of the escape time-scale $t_{\text{esc,ed}}(t)$ mimics the profile of the turbulent level. The acceleration time-scale profile is more complex due to its dependency on the inverse square of the varying Alfv\'en speed.

When dealing with two emission regions, one needs also to consider the possible contribution of external Compton (EC) processes: the flaring emission from the turbulent region is scattered off the relativistic electrons in the quiescent blob, and the steady-state emission from the blob off the relativistic electrons in the turbulent region. The former effect appears negligible for our conditions, but not the latter one. We thus use the sum of the synchrotron radiation of the flaring zone and of the quiescent blob as the seed photon field in the calculation of IC scattering on the electron population and of the IC cooling rate in the turbulent region. This leads to an average increase of flux in GeV-to-TeV $\gamma$-rays by $\sim 40$ per cent and a decrease of the flux in soft-to-hard X-rays by a similar value, with respect to the case where this EC effect is not included. The effect is therefore rather significant.

For the adjustment of the model to the data, we vary the five free parameters finally describing the flaring state, namely the magnetic field strength $B_{\text{tr}}$ and the typical size of the turbulent region $R_{\text{tr}}$, the longest wavelength in the wave-turbulent spectrum $\lambda_{\text{max}}$, regulating the longest escape time-scale during the turbulence, and the rise and decay time-scales of the turbulent energy injection rate defined in the source frame, $t_{\text{turb,r}}$ and $t_{\text{turb,d}}$ respectively. The sum of the emission of the quiescent blob and the time-dependent emission of the turbulent region are then compared to the observed MWL emission.

\subsubsection{Results}

\begin{figure*}  
\includegraphics[height=95mm]{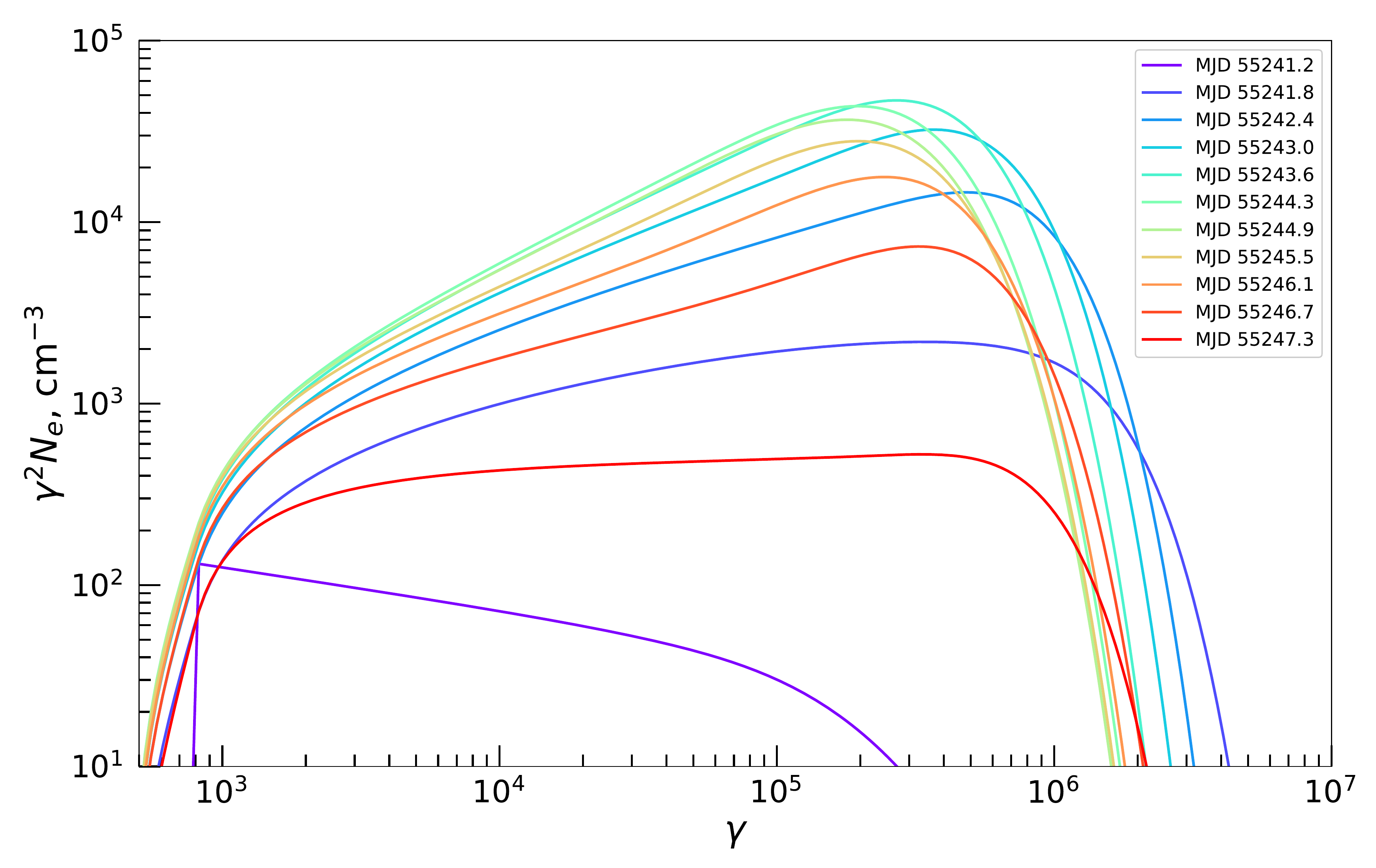}
\caption{Simulated time evolution of the electron spectrum in the turbulent region during the flare. Electrons are injected into the turbulent region with Lorentz factors above $\gamma_{\text{min,inj}} = 800$, as they are initially injected into the quiescent blob above this Lorentz factor. The electron distribution then evolves from violet to red curves. The evolution is displayed with a time step of $\sim 0.6$ d.} 
\label{fig:twoemitzonesne}
\end{figure*}

The temporal evolution of the electron spectrum in the turbulent region is illustrated in Fig.~\ref{fig:twoemitzonesne}. One clearly distinguishes the effect of particle acceleration, leading to an increase in the maximum electron energy and a hardening of the particle spectrum. Once acceleration becomes inefficient, particle cooling and escape result in the decrease of the maximum energy and in spectral softening.

\begin{figure*}  
\hspace*{0mm} \includegraphics[height=103mm]{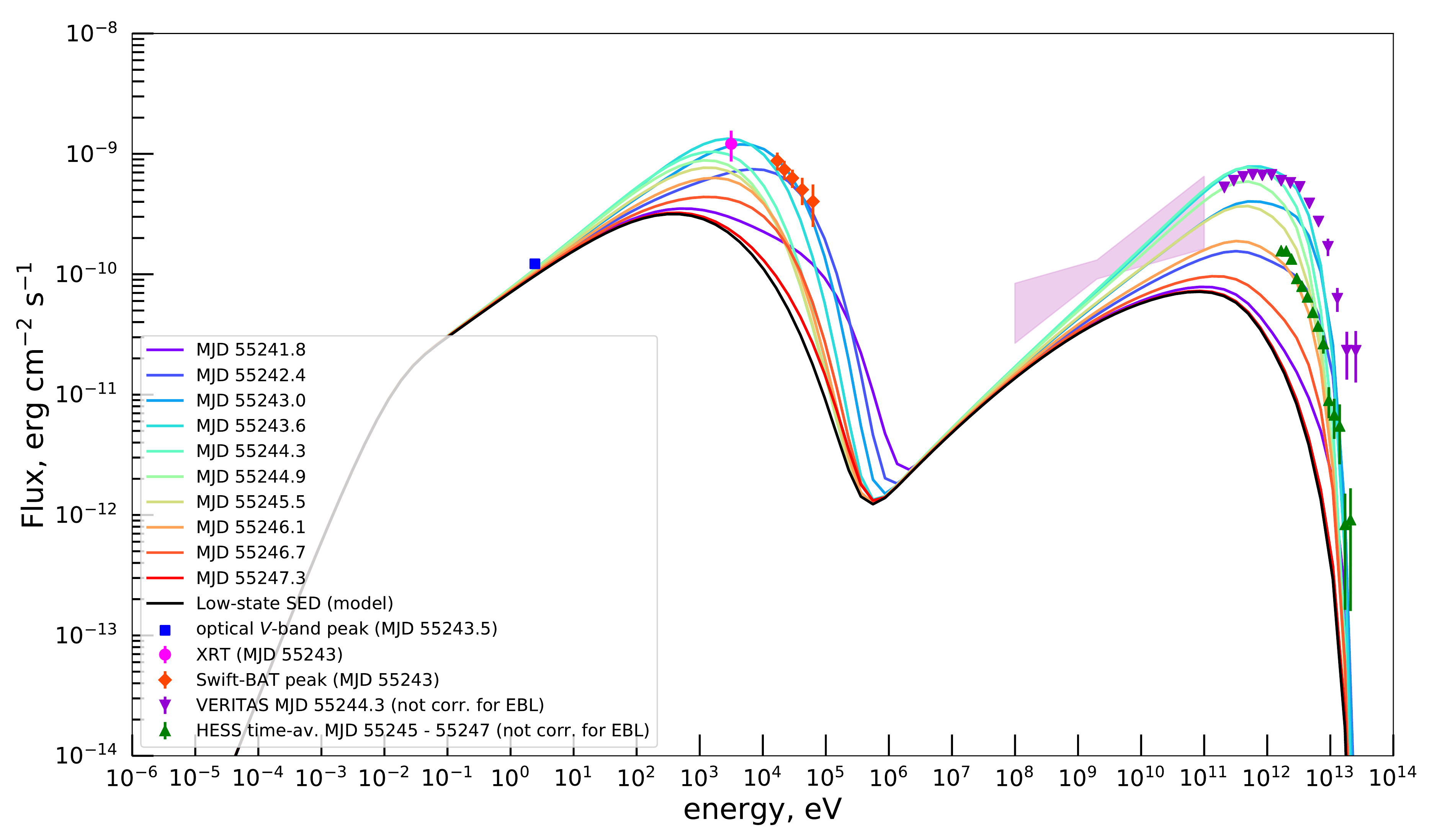}
\caption{Simulated time evolution of the broad-band SED during the February 2010 flare compared to the spectral measurements from the flare data set. The modelled SEDs are absorbed on the EBL using the EBL model by \protect\cite{dominguez}. The black line corresponds to the SED model of the low-state of the source, and the SED then evolves from violet to red curves during the flare. The evolution is shown with a time step of $\sim 0.6$ d. The blue square point indicates the optical flux during the flare peak (host galaxy subtracted), the magenta round point the flux detected by XRT at $\sim$3\,keV near the flare peak (16 February 2010, MJD\,55243), the red diamond points the {\it Swift}-BAT peak SED (16 February 2010), the violet down-pointing triangle points the VERITAS SED measurement during 17 February 2010 (MJD\,55244.3, not corrected for EBL), the green up-pointing triangle points the H.E.S.S.\ time-averaged SED over the period 17-20 February 2010 (MJD\,55245.0 -- 55247.0) during the flare decay (not corrected for EBL). The pink butterfly represents the {\it Fermi}-LAT SED uncertainty band for the spectral measurement during 16 February 2010. Optical data (host galaxy subtracted) is taken from \protect\cite{shukla2012}, VERITAS spectrum from \protect\cite{fortson}, H.E.S.S.\ measurements from \protect\cite{hessobsflare}. The data of XRT, {\it Swift}-BAT and {\it Fermi}-LAT are adapted from \protect\cite{singh2014}.} 
\label{fig:twoemitzonessed}
\end{figure*}

The corresponding time-dependent SED of the total emission obtained for our best solution is shown in Fig.~\ref{fig:twoemitzonessed}, with observational data superimposed. The overall agreement with the available spectra is very satisfactory. The two-zone model arrives at a good description of the flux increase in the X-ray and $\gamma$-ray bands, while the optical flux remains nearly unperturbed. Only the low-energy part of the {\it Fermi}-LAT spectrum during the flaring state appears slightly underestimated.

\begin{figure*}

\begin{minipage}[t]{.33\textwidth}
\begin{flushright}
\includegraphics[width=55.0mm]{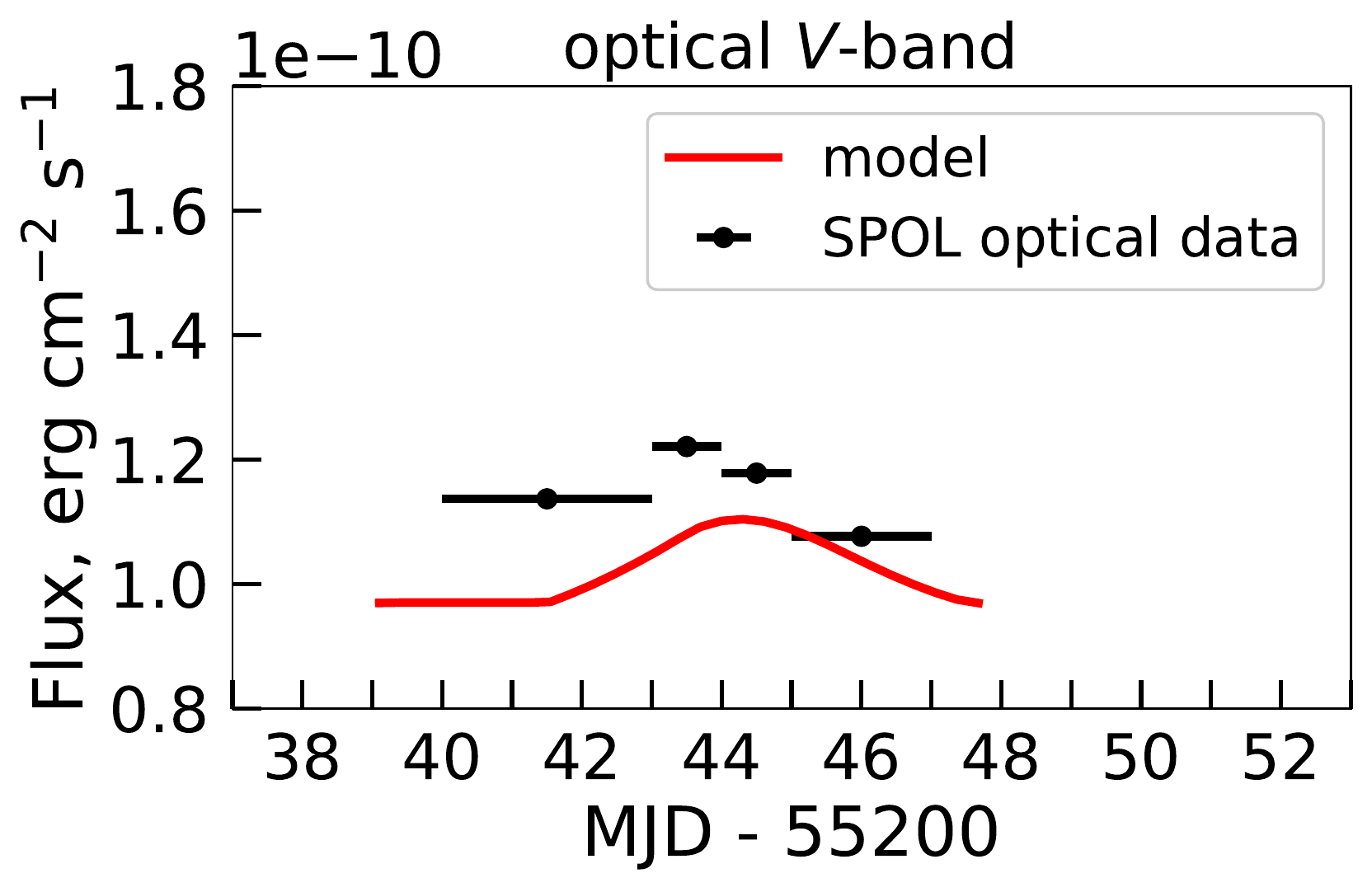}

\vspace*{1mm}

\includegraphics[width=52.6mm]{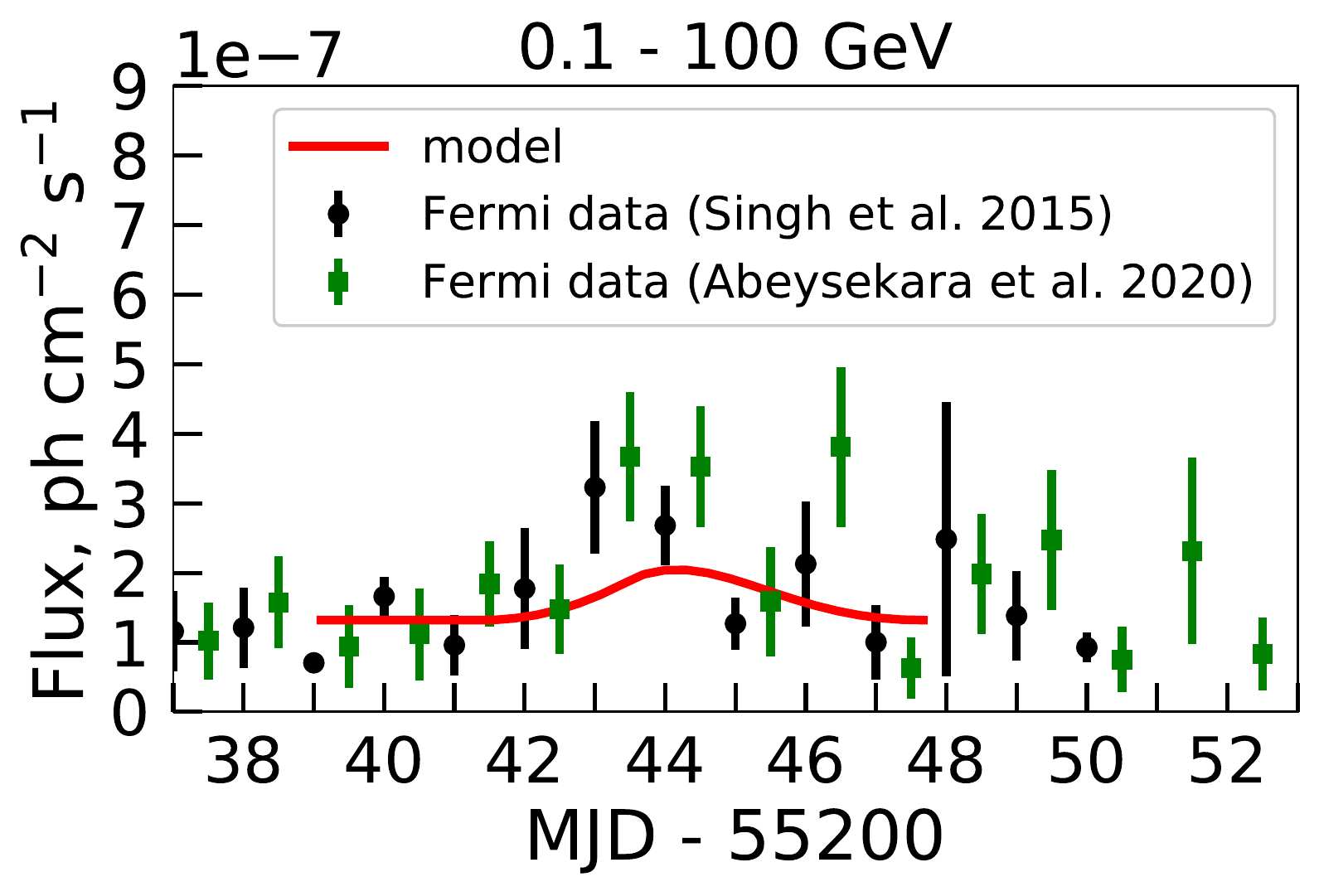}
\end{flushright}
\end{minipage}
\begin{minipage}[t]{.33\textwidth}
\begin{flushright}
\includegraphics[width=56.5mm]{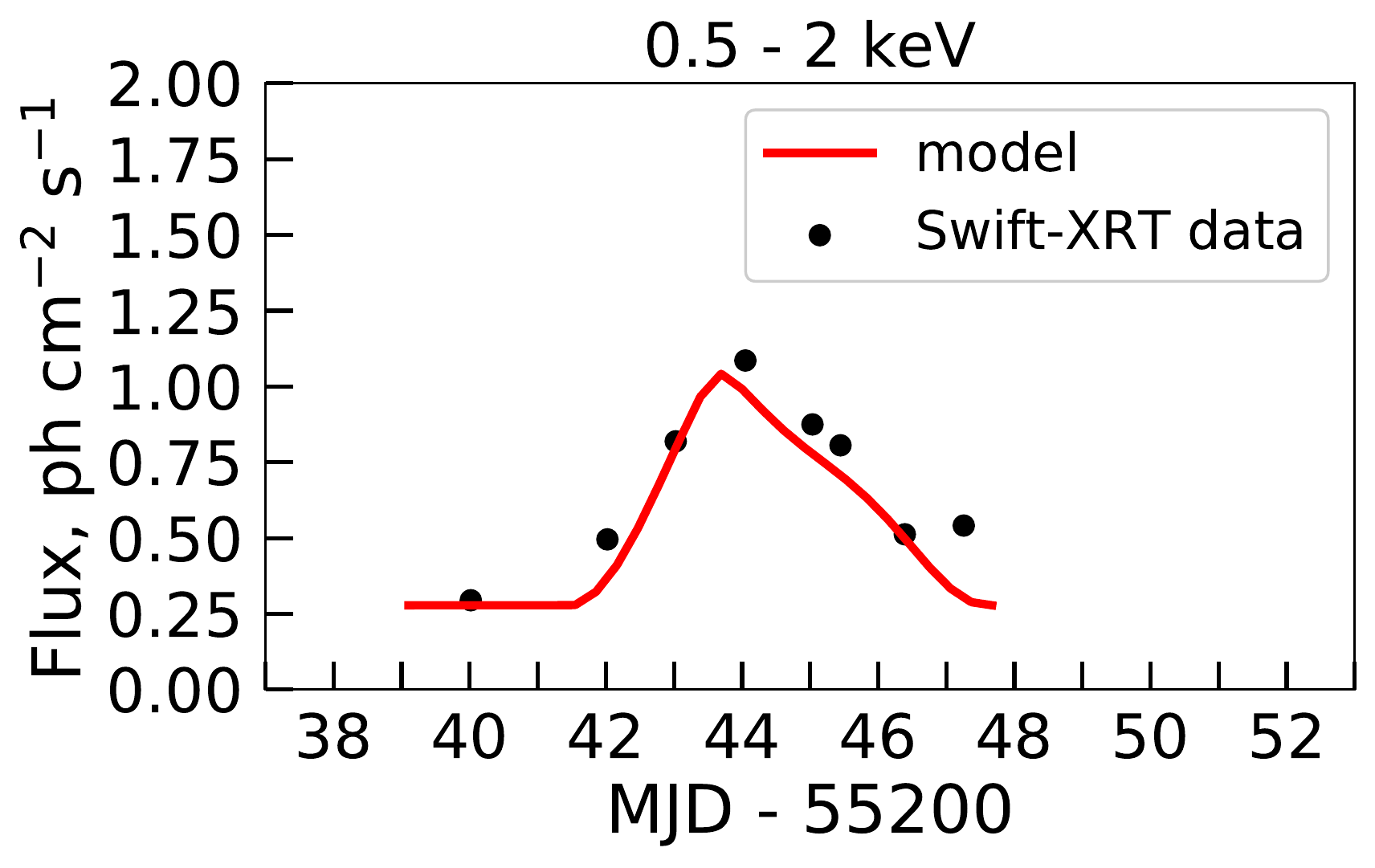}

\vspace*{1mm}

\includegraphics[width=52.6mm]{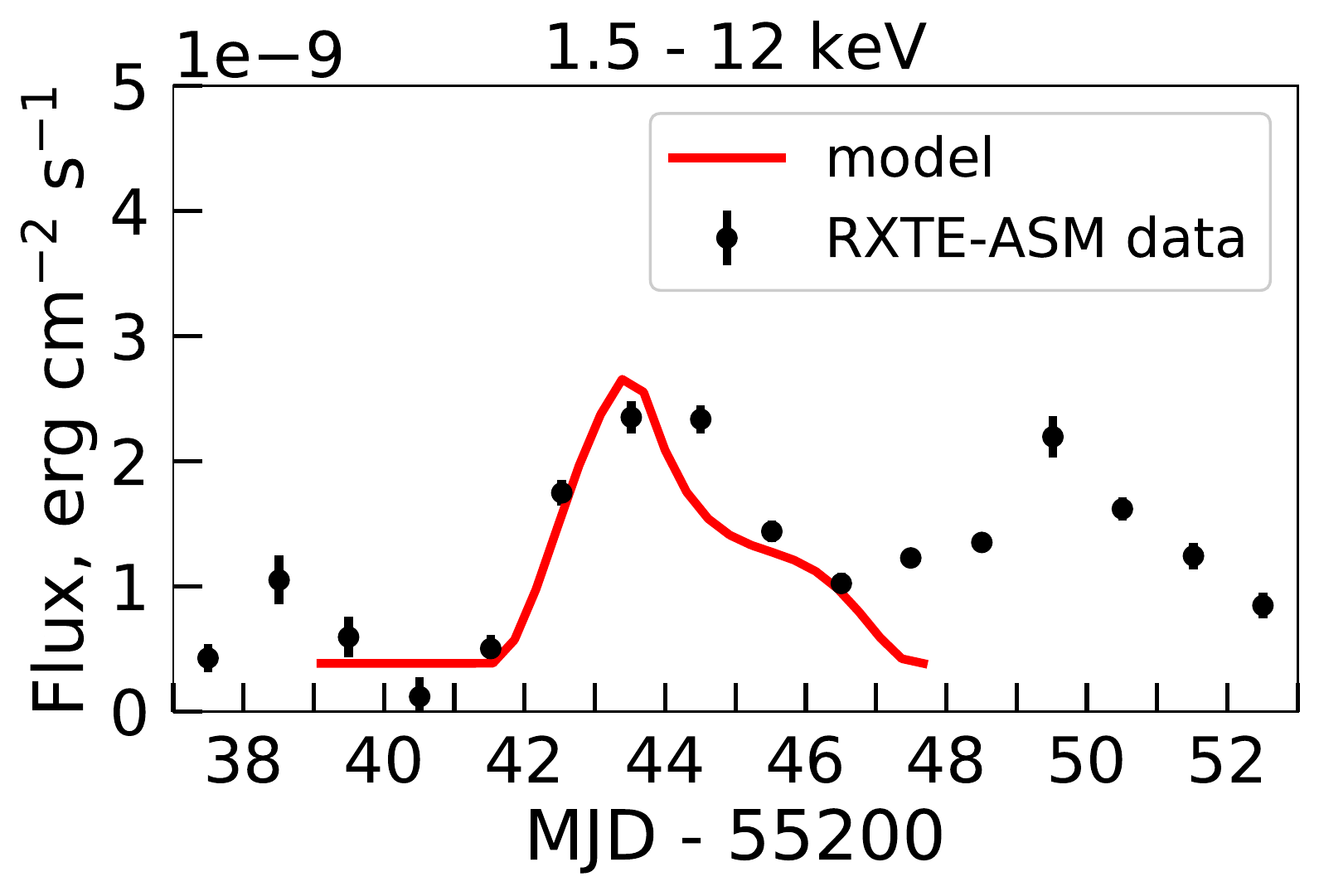}

\vspace*{1mm}

\includegraphics[width=55.0mm]{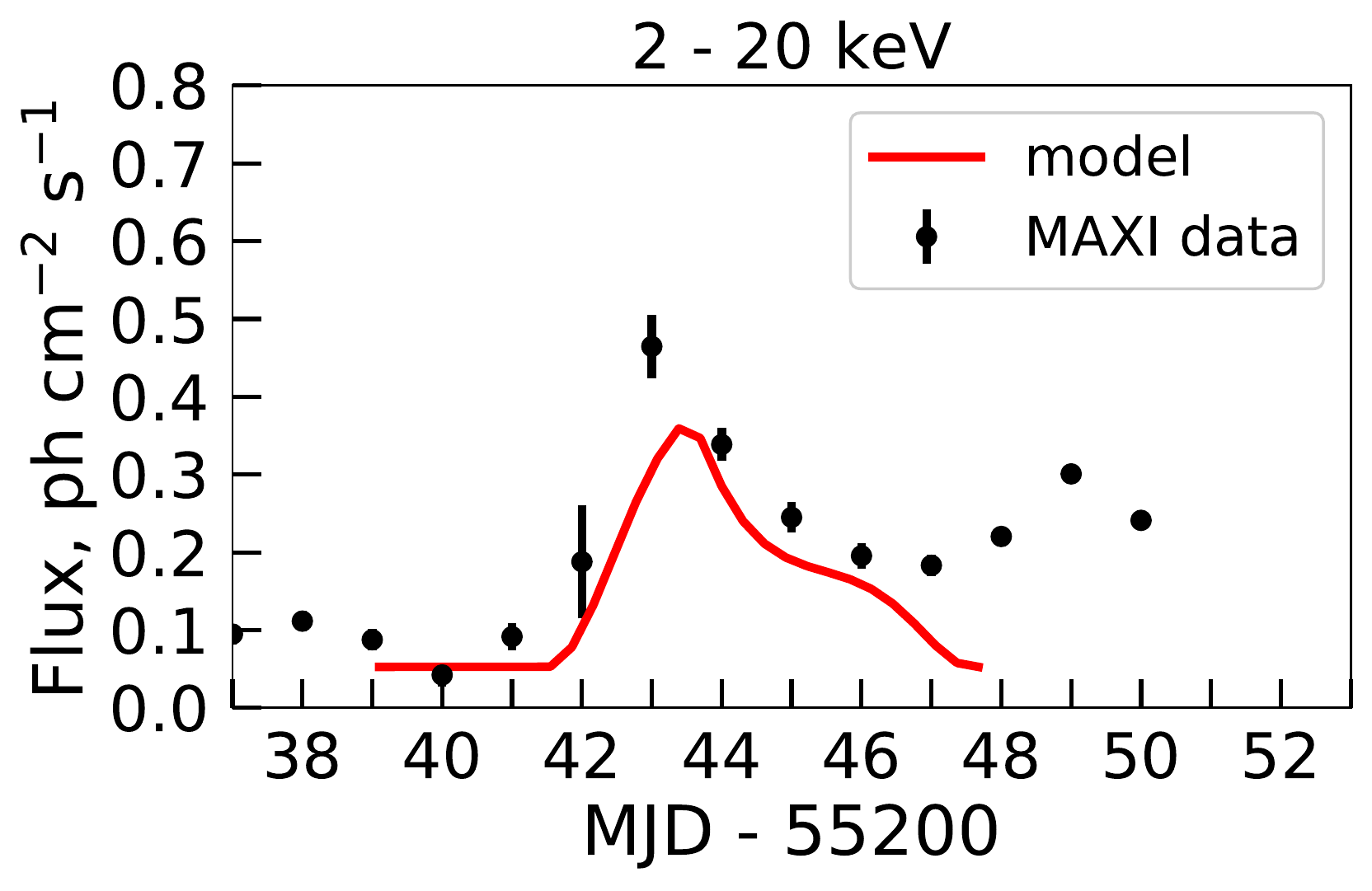}

\vspace*{1mm}

\includegraphics[width=55.0mm]{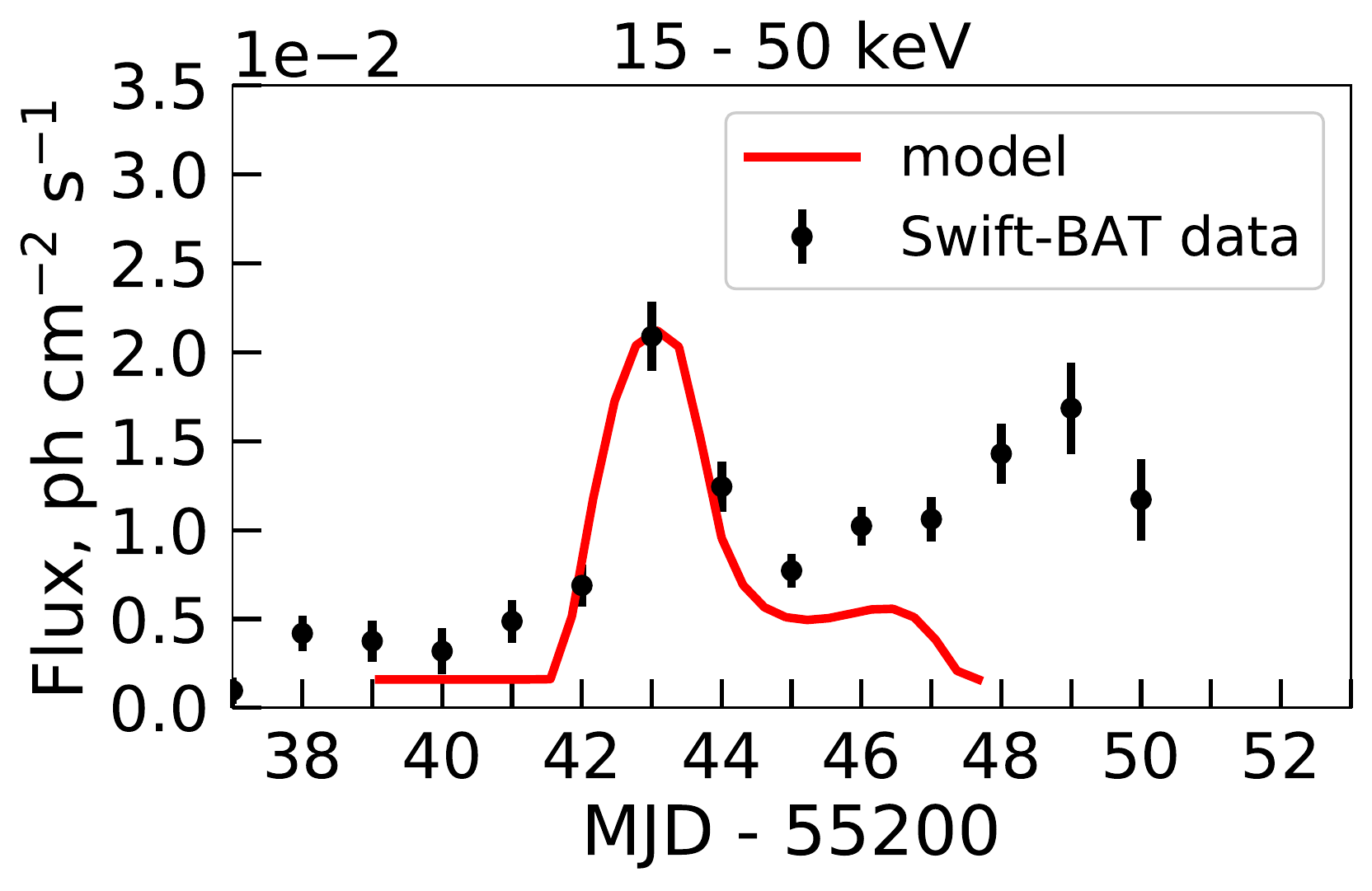}
\end{flushright}
\end{minipage}
\begin{minipage}[t]{.33\textwidth}
\begin{flushright}
\includegraphics[width=55.0mm]{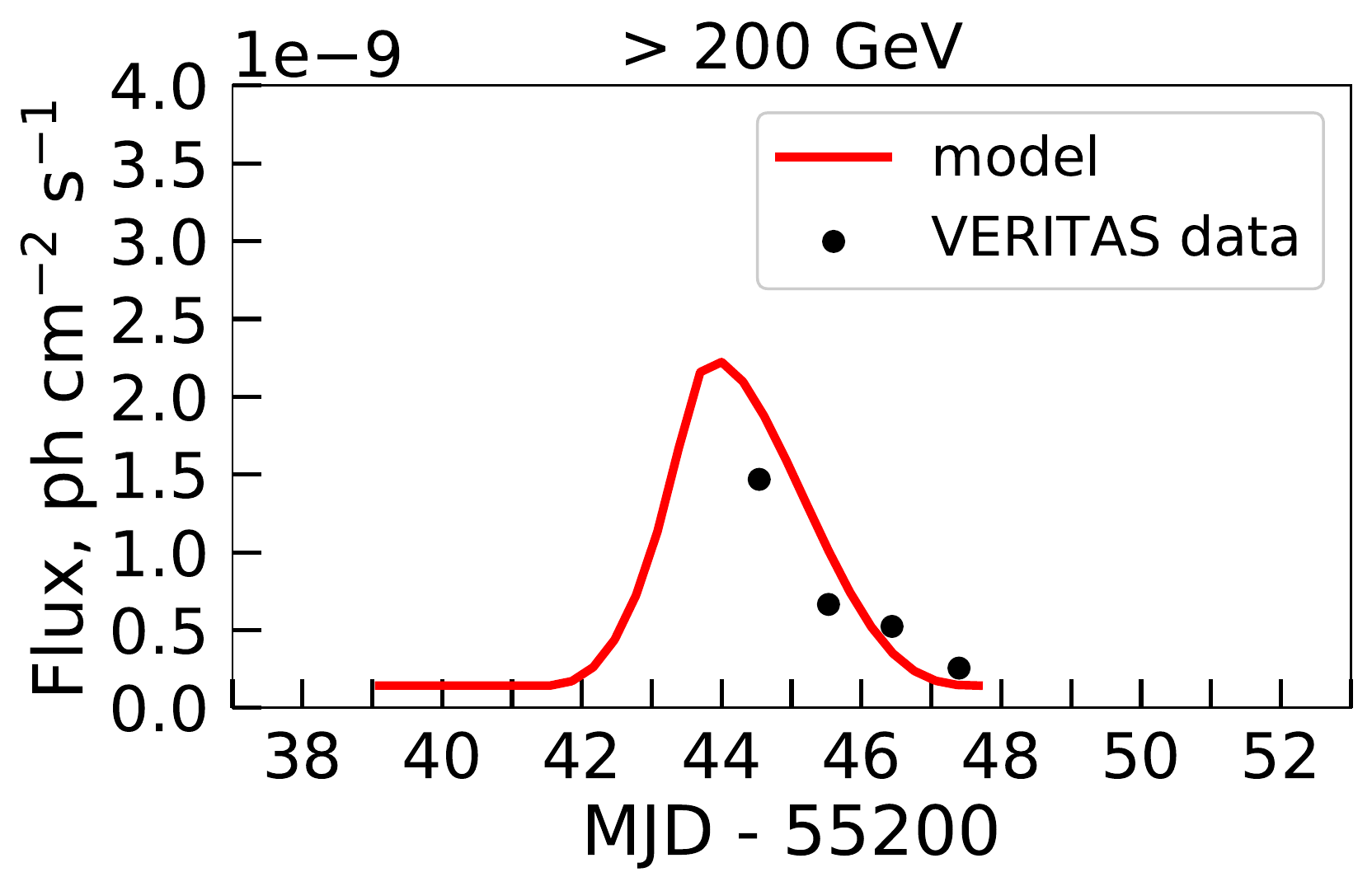}

\vspace*{1mm}

\includegraphics[width=55.0mm]{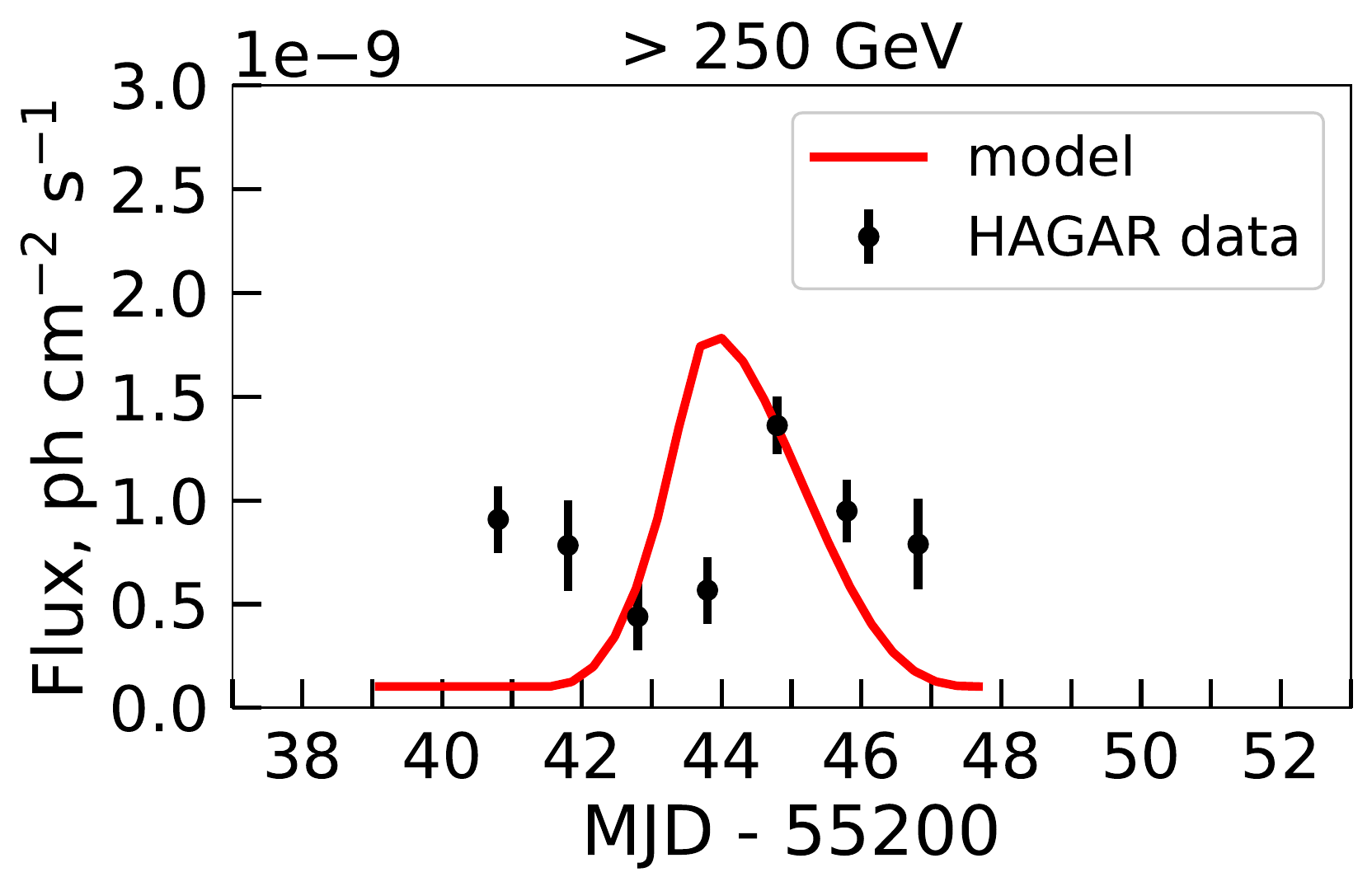}

\vspace*{1mm}

\includegraphics[width=55.0mm]{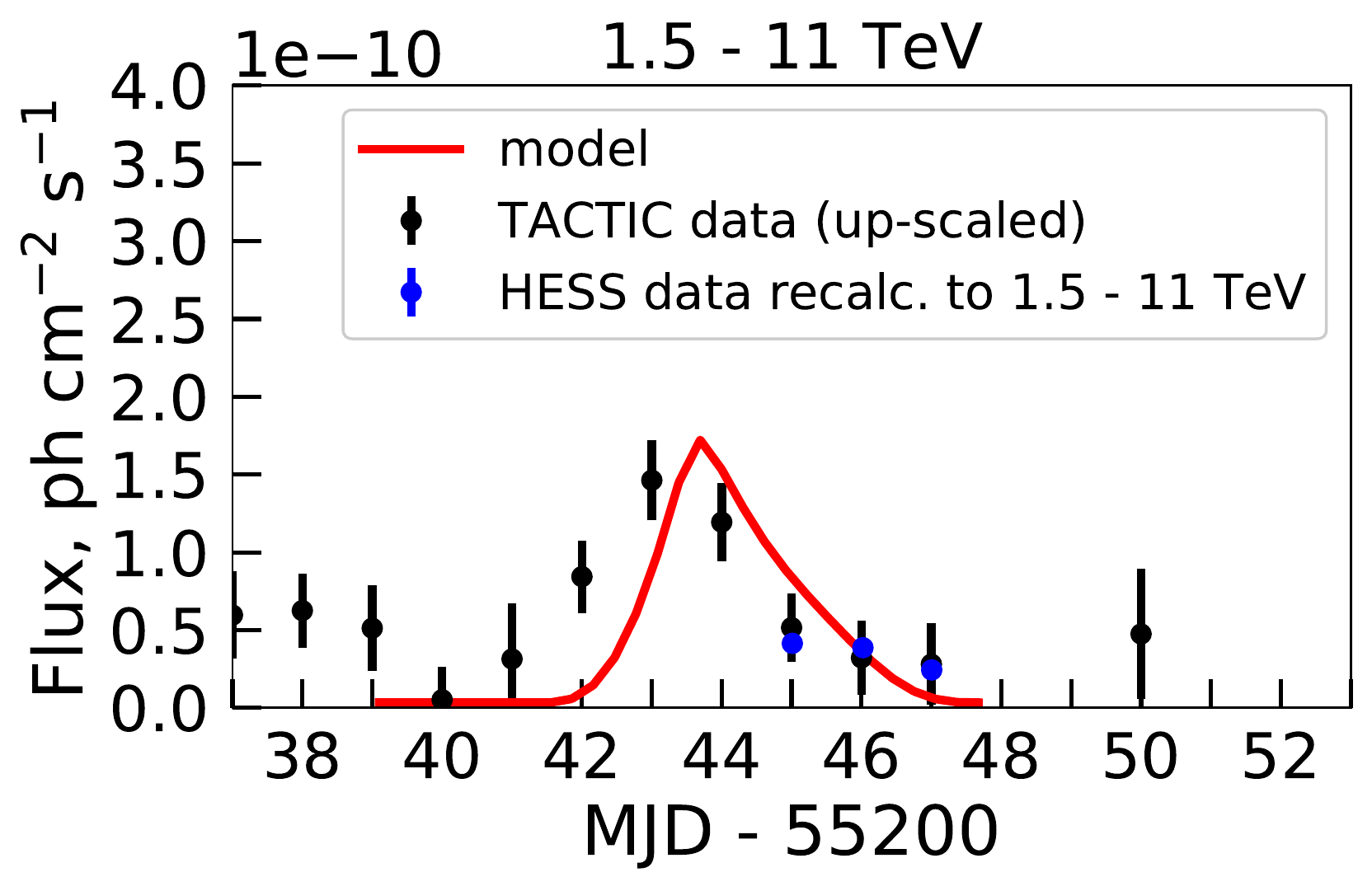}

\vspace*{1mm}

\includegraphics[width=56.55mm]{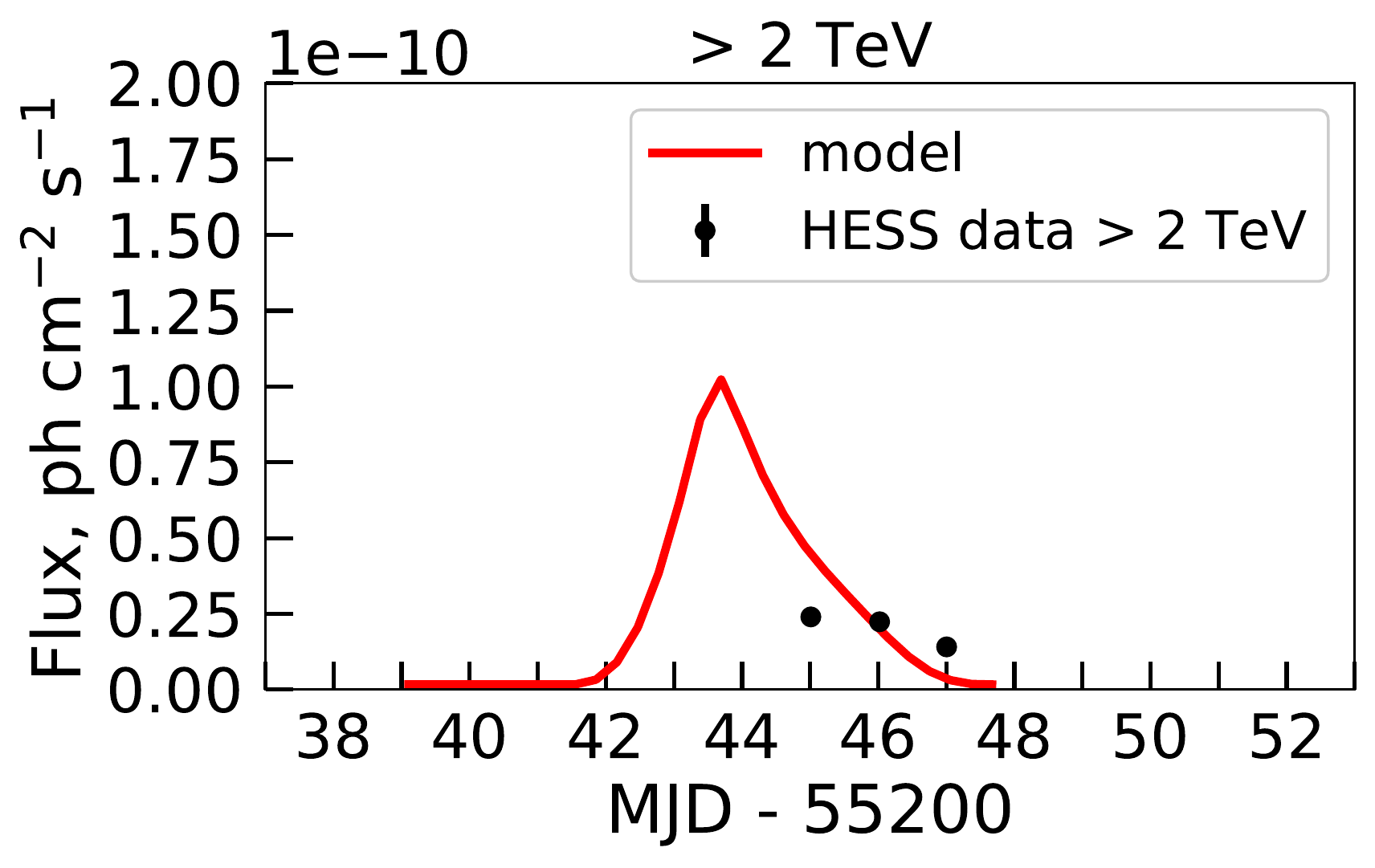}
\end{flushright}
\end{minipage}

\caption{Light curves generated by the two-zone model during the February 2010 flare compared to the MWL data. The observations include X-ray light curves by XRT and MAXI \protect\citep{singh2014}, {\it Swift}-XRT and RXTE-ASM \protect\citep{shukla2012}, the {\it Fermi}-LAT light curve computed by \protect\cite{singh2014} (with older instrument response functions) and by \protect\cite{veritasobsflare} (with newer instrument response functions), and the VHE light curves by H.E.S.S. \protect\citep{hessobsflare}, HAGAR \protect\citep{shukla2012}, TACTIC (recalibrated from \protect\cite{singh2014}), and VERITAS \protect\citep{veritasobsflare}. The optical flux time evolution (host galaxy subtracted) is taken from \protect\cite{shukla2012}.}
\label{fig:twoemitzoneslcs}
\end{figure*}

Fig.~\ref{fig:twoemitzoneslcs} shows the comparison between the simulated light curves and the MWL data in different energy bands. Ignoring the secondary peak that is visible in the X-ray band, the model provides a good representation of the observed flux variations, especially in the X-rays. The modelling  accurately reproduces the position of the X-ray peaks with the occurrence of the soft X-ray flux maximum about one day later than in the hard X-rays. The appearance of a `shoulder' following the peaks in the X-ray band is also a natural prediction of the turbulent acceleration scenario, as discussed below. In the optical band, while the flux variation has the right amplitude, there is a small offset between the observed and simulated flux amplitude at the level of $\sim 10$ per cent. However a systematic shift may be present (\textit{i}) in the optical data due to uncertainties in the host galaxy flux subtraction, and/or (\textit{ii}) in the modelling of the low-state net optical flux (note the non-negligible scatter of the data in the optical band in Fig.~\ref{fig:qsm}). Therefore, given that we reproduce the observed variability amplitude, our description of the optical data can be considered as acceptable.

\begin{table*}
\footnotesize
\centering
\begin{tabular}{ |c|c|c| } 
 \hline
 Parameters for the February 2010 flare & Symbol & Value \\ 
 \hline
 Magnetic Field [G] & $B_{\text{tr}}$ & $0.05$  \\ 
 \hline
 Comoving effective size of the turbulent zone [cm] & $R_{\text{tr}}$ & $3.65 \times 10^{15}$ \\
 \hline
 Maximal wavelength in the turbulent spectrum & $\lambda_{\text{max}}$ & $0.023 \, R_{\text{tr}}$ \\ 
 \hline
 Duration of the rise of the turbulent energy injection rate (source frame) [s] & $t_{\text{turb,r}}$ & $5 \times 10^6$ \\
 \hline
 Duration of the decay of the turbulent energy injection rate (source frame) [s] & $t_{\text{turb,d}}$ & $10^7$ \\
 \hline
\end{tabular}
\caption{Physical parameters describing the flaring state and the turbulent region in the two-zone scenario.}
\label{tab:twozonemodelbpar}
\end{table*}

Numerical values of the parameters describing the turbulent region are provided in Table~\ref{tab:twozonemodelbpar}. The effective size of the turbulent zone appears to be an order of magnitude smaller than the one of the quiescent blob, while the magnetic field in the turbulent region remains similar to the one in the quiescent blob. The maximal wavelength in the wave-turbulence spectrum cannot exceed the size of the turbulent region, $\lambda_{\text{max}} \leq R_{\text{tr}}$, since modes of the turbulent cascade cannot stretch beyond the turbulent region and  $\lambda_{\text{max}} > R_{\text{tr}}$ would imply superluminal escape of electrons (see Eq.~\ref{eq:tescedmf}). Our best-fitting model fulfils this condition with $\lambda_{\text{max}} \approx 0.023 \, R_{\text{tr}}$, well below the size of the turbulent region. We can also verify that the flux of particles from the turbulent region to the quiescent one remains negligible. The relevant injection rate (per unit of volume) can be estimated as 

\begin{equation}
    \label{eq:ejectionparticles}
    Q_{\text{inj,tr-qr}}(\gamma,t) \simeq \dfrac{N_{\text{e,tr}}(\gamma,t)}{2 \, t_{\text{esc,ed}}(t)} \, \left(\dfrac{R_{\text{tr}}}{R_\text{b}}\right)^3 
\end{equation}

assuming from geometrical arguments that half of the particles escaping the turbulent region reach the quiescent blob. We find that the spectrum of injected particles at the peak of the flare is an order of magnitude below the injection provided by the stationary front shock, therefore we conclude that the inflow of particles from the turbulent zone to the blob is indeed subdominant, and the emission from the quiescent blob remains approximately steady, in accordance with the proposed scenario.

The generic two-zone scenario reproduces in detail all the X-ray data related to the main flare (peaking at MJD\,55243 -- 55244), including its rise and decay temporal profiles, and the time lag of about 1\,d observed between hard and soft X-rays. The time-dependent injection of turbulence and in particular the gradual decrease of the energy input induce a characteristic feature in the light curves, namely a flux stagnation before the end of the flare that is visible in the X-ray band and, to a lesser degree, at higher energies. This occurs due to a faster escape of particles when the level of the turbulence drops significantly. Because of the faster electron escape, the energy density in the turbulent zone decreases, the Alfv\'en speed increases, competing with the decreasing turbulence level, which leads to a temporary stabilisation of the Fermi-II acceleration time-scale (see Eq.~\ref{eq:stochacctscaleturblevel}). The energy flux stalls, as the moderate acceleration balances cooling and escape. This continues until the fading of the turbulence becomes the dominant process and the stochastic acceleration time-scale quickly goes to infinity. Three X-ray light curves seem to show such a feature, before the secondary flare (peaking at MJD\,55246 -- 55247), although the presence of a second flare renders it difficult to establish. The model describes the behaviour of the X-ray energy flux, reaching the shoulder at that date. However the boost in Alfv\'en speed is not enough to initiate the observed secondary flare. Therefore, we argue that the secondary flaring event (which is seen only in X-rays, and not in $\gamma$-rays) might be caused by a `second wave' of weaker turbulent energy injection, with another rise and decline of the turbulence injection rate, which we do not model here. The required secondary peak of injection could be caused by e.g.\ a second dense cloud on the way, or an echo-like growth of another instability mode.

As already mentioned, the adopted two-zone model slightly underproduces the {\it Fermi}-LAT spectra. One way to better describe the observed 0.1 -- 1 GeV flux (Fig.~\ref{fig:twoemitzonessed}) without perturbing its synchrotron counterpart (optical flux) is to assume that there is an additional external Compton component. So far we have neglected the upscattering of flaring synchrotron emission off the quiescent blob electrons. With slightly different parameters of the turbulent region, this process could provide a more important contribution to the total GeV flux. In particular, a larger size of the turbulent zone would lead to a higher radiation density seen in the quiescent blob. In this case the particle flux from the flaring zone to the quiescent emission region might become rather important and has to be taken into account. The {\it Fermi}-LAT light curve does not provide a tight constraint on the model.

At VHE, the two-zone model does not reproduce the exact shape of the VERITAS and H.E.S.S.\ light curves, which are the most constraining, but it appears compatible with the data set. The model fits reasonably well the VERITAS spectrum obtained at MJD\,55244.3 (roughly 1 d after the flare peak), as well as the spectral measurement by H.E.S.S.\ averaged over time period MJD\,55245.0 -- 55247.0 (fall of the flare). Some discrepancy can be seen at the energies above $\sim 5$ TeV. While our modelled SED may be slightly too soft at the highest energies, due to a relatively simple treatment of the turbulent acceleration and possibly systematics of the EBL model, it should be noted that the VERITAS spectrum points are still preliminary and may suffer from systematic uncertainties. Focusing on the night-to-night variations, we did not try to treat the intra-night variability. One may consider that the fluctuations of the flux at the 1\,h time-scale can be due to small flaring subregions in the emitting turbulent region. Indeed, we did not model the spatial structure of the turbulence region and approximated it for practical reasons with four independent spheres, while in reality it might be composed of many individual small cells having random direction of magnetic field and fluctuating velocity fields and electron density \citep{marscher2014}. Such more complex physics of the turbulent region could explain stochastic flux variations at short intra-night time-scales.

\section{Discussion}
\label{sec:discussion}

The generic two-zone scenario developed in this paper provides one of the first fully time-dependent models able to reproduce reasonably well the evolution of the MWL flare despite a limited number of free parameters (Table~\ref{tab:lowstateparams} and \ref{tab:twozonemodelbpar}), relying on the interpretation of the flare as a weak perturbation of the quiescent phase. The time evolution of the flare is dominated by two key parameters, the rising time of the turbulent energy injection rate $t_{\text{turb,r}}$, and the effective size of the turbulent zone $R_{\text{tr}}$. However, the uniqueness of the selected (imperfect) solution can not be ensured.

\subsection{The assumption of flare as a weak perturbation}

Our initial hypothesis describing VHE flares as weak perturbations of the quiescent state appeared quite powerful to significantly constrain the flare models and to reduce the number of free parameters, but it should be tested by applying it to other flares once more detailed and complete MWL monitoring data sets become available. A preliminary analysis shows that the 2008 flare of Mrk\,421 could be another example \citep{donnarumma2009}. Indeed, AGN flares are just non-destructive transient events and, although remarkable during the outbursts, their total energy budget remains negligible compared to the energy radiated over years by the quiescent states. However, a limitation to this view could arise from a detection of extended VHE emission in blazars, unless most of the quiescent emission still comes from a compact radiatively dominant region. So far, extended VHE emission in AGN was detected only from the jet of a non-blazar source, the radio galaxy Cen\,A \citep{cenahess2020}. This result questions the validity of the present scenario for radio galaxies, which should be further investigated if VHE flares were detected from Cen\,A, which is not yet the case.

\subsection{Alternative models}

The one-zone model has served the SED modelling well in the past. However the situation becomes more challenging with increasing quality and completeness of data sets. One-zone scenarios for the February 2010 flare of Mrk\,421 did not prove to be successful under our assumptions and several alternative two-zone scenarios can be considered. We developed here the two-zone scheme which appeared to us the most promising, based on previous works available in the literature on the February 2010 flare, but other options could be further analysed as well. As mentioned before, an alternative two-zone scenario can be provided by a non-radiative acceleration region around the emitting blob \citep{kirkmastichiadis}, which injects particles, but contributes only a negligible amount of radiation due to a low magnetic field. The acceleration region would be injecting particles with a hard spectrum in the quiescent blob to launch the flare. Such a scenario should in principle be able to provide the inflow of electrons with a hard spectrum that is needed to arrive at a strong flux increase at high energies, while the optical flux remains relatively constant. However, as shown in \cite{dmytriievicrc} this scenario describes indeed well the varying synchrotron emission, but it underpredicts the $\gamma$-ray flux by a factor of $\sim 3$ for the data set under study. A two-zone model considered by \cite{caowang} in their modelling of the June 2008 flare of Mrk\,421 assumes that the steady emission arises from the outer jet where particles are accelerated by the Fermi-I process and dominates the total flux in the radio-to-optical domain, while the variable component is produced in a much smaller inner jet region in which low-energy particles are injected continuously and undergo stochastic acceleration. The flare is then initiated by a change in the stochastic acceleration time-scale. Such a configuration is interesting, but the model still lacks a full time-dependent framework to self-consistently describe the SED evolution from the low state to the high one.

\subsection{Complexity of the physics involved in the scenario}

The turbulent acceleration of particles as the process driving flares of HBLs was previously considered by various authors. Different types of turbulence can be assumed. For instance, \cite{tramacere} reproduce various trends observed in six HBL flares with stochastic acceleration of particles on a relatively short time-scales and conclude that two acceleration scenarios can provide a good description of the X-ray data: (\textit{i}) the Fermi-II acceleration time-scale evolves because of variations of $\delta B^2$ or $\beta_\text{A}$, while the `hard-sphere' turbulence spectrum remains constant, or (\textit{ii}) the index of the turbulence spectrum evolves with time. In the present paper, we stick to the first scenario, considering however a more complex self-consistent evolution of the energy density of magnetic field fluctuations and of the Alfv\'en speed, as well as following a two-zone approach.

Several effects can lead to the generation of a turbulence in the vicinity of the emitting quiescent blob, assumed to stream at relativistic speed along the jet axis. One possibility is a sudden enhancement of plasma instabilities, or Kelvin-Helmholtz or rotationally-induced Rayleigh-Taylor instability \citep{meliani} in the shear at the interface between the quiescent blob associated to a faster inner spine and the slower outer layer of the jet \citep{sol1989}. The turbulence might also be excited as the blob is passing through a dense cloud of gas, resulting from e.g.\ an interaction between a red giant star and the jet \citep{barkov2012}. Due to enhanced density of the plasma flowing past the blob, the Reynolds number could appear to be higher than the critical one, and as a result, a transient turbulent region is formed. The exact physical conditions and plasma instabilities that trigger spontaneous and transient turbulence around the emitting blob question the broad field of turbulence generation and shock-turbulence interaction, and cannot be simulated in simple radiative models as developed here. As described above, we approximate the evolving injection rate of the turbulent energy with a simple linear rise and decay. However, we find that the exact shape of the injection function has only a small impact on the resulting shape of the light curves, as long as it represents a peak, characterised by suitable rise and fall time-scales. So, finally, a very rough modelling of the turbulence can still provide a reasonably good description of the observed event. 

Clearly, all current available flare scenarios need to proceed with highly simplifying assumptions in view of the limited current knowledge on the VHE emitting regions, and of the expected underlying complexity of the physics involved. For instance in the two-zone scenario developed here, one should probably expect some deformation of the `stationary' front shock as a back reaction to the external perturbation supposed to induce the turbulence and the flaring at the edges of the quiescent blob, as well as a possible mutual shock-turbulence interaction \citep{andreop}. A large number of linear and non-linear phenomena may occur which can potentially deform the stationary shock, impact on the quiescent emission, and modify the characteristics of the turbulence, which we completely neglected here. Instead of injecting directly the turbulence at the flanks of the central blob as a boundary effect, another possibility that we did not take into account could be as well to consider that the external perturbation acts first on the `quiescent' front shock, which in turn can enhance the turbulence behind it and around the central blob. Magnetic reconnection could also play a role in such collisionless plasmas (\cite{karimabadi} ; \cite{nishikawa2020}). However this process is not expected to contribute significantly to particle acceleration in our scenario involving a weakly magnetised shock.

\subsection{A new tentative interpretation of lognormality and different types of noises in blazars}

In the two-zone scenario developed here, turbulence effects underlie both the quiescent emission and the flaring one, in a slightly different way through Fermi-I and Fermi-II mechanisms. In such a framework, it seems possible to attribute the lognormality observed in VHE light curves of some bright AGN, such as PKS\,2155-304 (\cite{aharonian2007}; \cite{abram2010} ; \cite{abdalla2017}), to a turbulent process somewhat similar to the universal one suspected in laboratory turbulent flows \citep{mouri}, and to which the multiplicative central-limit theorem can be applied. Indeed, particle-in-cell simulations have also put in evidence a lognormal distribution of the particle and internal energy density in turbulent collisionless, magnetised, relativistic electron-positron plasmas similar to those  expected in some AGN jets \citep{zhdankin}. Under such interpretation, detailed studies of the power spectral density (PSD) could help characterising the turbulence properties directly in the VHE emitting regions.

The measured PSDs appear as a power law $P(\nu_{t}) \propto \nu_{t}^{-\beta}$ where $\nu_{t}$ is the temporal frequency, with a variable index $\beta$ in the range $1 \leq \beta \leq 2$. This reveals how the amplitudes of the variability are spread over the different time-scales and suggests that stochastic underlying processes are at work in the emitting regions, with correlated coloured-noises typically of the flicker or pink type ($\beta = 1$), or of the random walk or red type ($\beta = 2$). Different values of the $\beta$ index have been found for the quiescent and the flaring states in the two blazars PKS\,2155-304 and in Mrk\,421, with the same trends. They suggest a pink noise for the quiescent VHE states and a red noise for the VHE flaring states, with $\beta = 1.1^{+0.10}_{-0.13}$ on time-scales larger than one day to several years for PKS\,2155-304 \citep{abdalla2017} and $\beta = 1.1^{+0.5}_{-0.5}$ for Mrk\,421 on time-scales from months to years \citep{goyal}, and with $\beta \simeq 2$ for PKS\,2155-304 on time-scales from minutes to a few hours \citep{aharonian2007} and $\beta \simeq 1.75$ for Mrk\,421 on time-scales from seconds to hours during the night of the flare on February 17, 2010 \citep{veritasobsflare}. In both sources the PSD is flatter for quiescent states, and steeper for flares. A possible interpretation of these results, in the frame of the present two-zone model, suggests that the different acceleration mechanisms and underlying turbulences during the quiescent state or during the flares are at the origin of the different types of noises currently observed. Slowly variable shock acceleration (and particle injection) by the shock at the front of the blob, and long-term slowly varying turbulence inside the blob (with $q \leq 2$), allow a better spread of the amplitudes of the variability over the different time-scales above 1 day, considering that long-term perturbations of the front shock can occur on time-scales of days, months and years. Conversely, transient Fermi-II acceleration with $q=2$, for the `hard-sphere' turbulence, likely stores larger amplitudes of the variability at the time-scales of hours (corresponding to the largest spatial scales of the entire turbulent zone) compared to the smallest time-scales (minutes or seconds) due to turbulent cascade phenomena, resulting in a redder noise than for the quiescent long-term emission.

\subsection{Analogy with hotspots of extragalactic radio sources}

As a final remark, we would like to emphasise that the complex configuration adopted for the generic scenario with two zones and both Fermi-I and Fermi-II mechanisms, was just reached step by step, starting from the simplest scenario and adding complexity only when needed in view of the constraints imposed by the data. Unexpectedly, such a circumstance has some similarities with the situation observed in hotspots of nearby radio galaxies like 3C\,445, 3C\,105, 3C\,227, and 3C\,195, for which different authors advocate multi-zone models for particle acceleration, involving both Fermi-I and Fermi-II mechanisms to describe the hotspot physics with a compact front shock and a more diffuse and turbulent region in its wake (\cite{krulls} ;  \cite{prieto} ; \cite{fan}; \cite{orienti2012} ; \cite{isobenaoki} ; \cite{orienti2017} ; \cite{migliori2020} ; \cite{orienti2020}). So, finally, the scenario built independently in this paper for the complex quiescent and flaring VHE emitting zone of Mrk\,421 describes it as a kind of `mini hotspot', appearing much earlier along the jet. Although compactness, energies and temporal evolution involved are quite different, this analogy could provide some clues to better describe the VHE zone from information gathered at the larger and resolved scale of the hotspots.

\section{Conclusions}
\label{sec:conclusion}

We have provided a general analytical approach to determine the feasibility of a one-zone shock model to produce flaring events in blazars.

It was shown that the MWL data set of the February 2010 flare of Mrk\,421 cannot be described with a one-zone shock model, and neither with turbulent acceleration in a one-zone model.

A self-consistent two-zone model, with a large emission region responsible for the steady-state emission and a smaller, connected turbulent region responsible for particle acceleration and emission during the flare, provides a very satisfactory description of the available MWL spectra and light curves for this event. The observed spectral hardening and asymmetric flare profile are a direct outcome of the simulated acceleration, cooling and particle escape processes.

In this model, the steady-state emission and flare emission can be connected with a limited number of free parameters, and the flare arises naturally as a perturbation due to a transient turbulence on the edge of the steady-state emission region.

In general, the scenario we present is a viable model for producing flares on a day time-scale. A future application to further flare data sets should show whether the proposed intermittent turbulent acceleration provides an interesting scheme to explain most blazar flares.

\section*{Acknowledgements}
The authors wish to thank M.~Lemoine for useful discussions, C.~Nigro for his permission to share his numerical implementation of the Chang and Cooper scheme (without acceleration), M.~Meyer for the public availability of his python code to model a variety of EBL models. We also acknowledge S.~Buriak for assistance in designing the sketches representing the flare scenarios. Finally, we thank J.~Finke for providing us the data points of the Mrk\,421 low state, K.~K.~Singh for providing the data points of the spectral measurements in different energy bands and MWL light curves during the Mrk\,421 February 2010 flare (presented in his 2014 paper) and M.~Tluczykont for the H.E.S.S.\ data points measured during the flare.

\section*{Data availability}
The data sets used in this paper were derived from articles in the public domain (references presented in Section~\ref{sec:data}), and will be shared on reasonable request to the corresponding author. The numerical EMBLEM code used for the data modelling cannot be shared publicly due to its ongoing development and extension. More information on the code can be obtained on reasonable request to the corresponding author.

%%%%%%%%%%%%%%%%%%%%%%%%%%%%%%%%%%%%%%%%%%%%%%%%%%



%\bibliographystyle{mnras}

\bibliographystyle{mnras}
\bibliography{bibliography}

%\begin{thebibliography}{99}
%\bibitem[Abdo et al. (2011)]{abdo2011}
%Abdo, A. A. et al., \emph{Fermi Large Area Telescope Observations of %Markarian 421:  The Missing Piece of its Spectral Energy Distribution}, ApJ, %736:131. (2011)
%\end{thebibliography}

%%%%%%%%%%%%%%%%%%%%%%%%%%%%%%%%%%%%%%%%%%%%%%%%%%

%%%%%%%%%%%%%%%%% APPENDICES %%%%%%%%%%%%%%%%%%%%%

\appendix
%\section{more detailed description of the code ?  - if needed}

\section{Analytical solution of the kinetic equation for the case of shock perturbing a steady-state electron spectrum}
\label{appen:analyticalsolutionke}

Here we solve the Eq.~\ref{eq:kineqshockaccel} describing perturbation of electron population in the VHE emitting zone by a transient shock. Two key parameters of the passing shock govern the evolution of the electron spectrum $N_{\text{e,FI}}(\gamma,t)$ and hence of the MWL emission: the shock acceleration time-scale, $t_{_{\text{FI}}}$, characterising the efficiency in acceleration of particles, and $t_{_{\text{dur,FI}}} = t_{\text{cs}}$, which is a duration of the shock acceleration activity in the blob, equal to the passage time of the shock through it (in the frame of the blob). The latter parameter is linked to the time-scale of the flux rise in the light curve profile of the flare.

\subsection{Assumptions and boundary conditions}

We assume that the physical parameters of the emitting zone do not change during the passage of the shock. The shock enters the blob at $t = 0$.

The initial condition is that the electron spectrum at $t = 0$ is the steady-state solution $N_{\text{e},0}(\gamma)$:

\begin{equation} \label{eq:initcond1}
   N_{\text{e,FI}}(\gamma,t=0) = N_{\text{e},0}(\gamma) 
\end{equation}

The $N_{\text{e},0}(\gamma)$ is the asymptotic stationary solution of the kinetic equation with only injection, escape and cooling terms, and no acceleration process, deduced from the Eq.~\ref{eq:kineqshockaccel} (setting shock acceleration term to zero, $a = 0$):

\begin{equation} \label{eq:kineqsteady}
 		\dfrac{\partial}{\partial \gamma}\left( b_\text{c}\gamma^2 \cdot N_{\text{e},0}(\gamma)\right) - \dfrac{N_{\text{e},0}(\gamma)}{t_{\text{esc}}} + Q_{\text{inj}}(\gamma) = 0
 \end{equation}

We neglect the inverse Compton cooling, so the $b_\text{c}$ is constant in time. We also assume $t_{\text{esc}}$ is constant in time and energy-independent: $t_{\text{esc}} \sim R_\text{b}/c$, where $R_\text{b}$ is size of the emitting region. We require a boundary condition such that the electron spectrum tends to zero at the maximal Lorentz factor $\gamma_{\text{max}}$: $N_{\text{e},0}(\gamma=\gamma_{\text{max}}) = 0$. With this condition, this equation has the following solution:

\begin{equation} \label{eq:steadystatesolut}
    N_{\text{e},0}(\gamma) = \dfrac{1}{b_\text{c} \gamma^2} \int_{\gamma}^{\gamma_{\text{max}}} Q_{\text{inj}}(\gamma^{\prime}) \cdot \text{exp} \left( \frac{1/\gamma^{\prime} - 1/\gamma}{b_\text{c} t_{\text{esc}}} \right) d\gamma^{\prime}
\end{equation}

The multiplicative term in this expression and the negative exponent in the exponential describe how the injection effect is respectively damped by the cooling and by the escape. As already discussed, in this paper, we try to not restrict too much the evolution of the particle distribution by artificial boundaries, so we set $\gamma_{\text{max}} \rightarrow \infty$ when evaluating the steady-state electron spectrum given by the Eq.~\ref{eq:steadystatesolut}. Now let us consider how this spectrum is modified with time when the shock acceleration is acting on this electron population. Let us decompose the electron spectrum in the emitting zone during the passage of the shock into the initial and perturbed parts:

\begin{equation} \label{eq:decompose}
  N_{\text{e,FI}}(\gamma,t) = N_{\text{e},0}(\gamma) + N_{\text{e,p}}(\gamma,t)  
\end{equation}

The initial electron spectrum is the steady-state solution, and the time-dependent perturbed part is the one causing the flux increase. We plug this expression into the kinetic equation (Eq.~\ref{eq:kineqshockaccel}), which yields:

\begin{multline*} 
 		\dfrac{\partial N_{\text{e,p}}(\gamma,t)}{\partial t} =  \dfrac{\partial}{\partial \gamma}\left( W(\gamma) \, N_{\text{e,p}}(\gamma,t)\right) - \dfrac{N_{\text{e,p}}(\gamma,t)}{t_{\text{esc}}} - \dfrac{\partial}{\partial \gamma}\left(\dfrac{\gamma}{t_{_{\text{FI}}}} \, N_{\text{e},0}(\gamma)\right) 
 		\end{multline*}

 \vspace*{2mm}
 		
where $W(\gamma) = b_\text{c}\gamma^2 - \frac{\gamma}{t_{_{\text{FI}}}}$. \\

This equation describes the time evolution of the perturbed time-dependent addition $N_{\text{e,p}}(\gamma,t)$ to the steady state solution. Here $t_{_{\text{FI}}}$ is assumed constant in time and energy-independent. Plugging the expression for $N_{\text{e},0}(\gamma)$ (Eq.~\ref{eq:steadystatesolut}) to the equation, and evaluating the last term (free term depending only on $\gamma$), we find:

\begin{multline} \label{eq:kineqforperturb}
 		\dfrac{\partial N_{\text{e,p}}(\gamma,t)}{\partial t} =  \dfrac{\partial}{\partial \gamma}\left(W(\gamma) \, N_{\text{e,p}}(\gamma,t)\right) - \dfrac{N_{\text{e,p}}(\gamma,t)}{t_{\text{esc}}} \, + \, \mathcal{F}(\gamma) 
 		\end{multline}
 		
 	\vspace*{2mm}

 with $\mathcal{F}(\gamma) =  \left[ 1 - \frac{1}{b_\text{c} \gamma \, t_{\text{esc}}} \right] \, \dfrac{N_{\text{e},0}(\gamma)}{t_{_{\text{FI}}}} \, + \, \dfrac{Q_{\text{inj}}(\gamma)}{b_\text{c} \gamma \, t_{_{\text{FI}}}}$. \\ 

We obtained the final form of the equation governing how $N_{\text{e,p}}(\gamma,t)$ is evolving with time. The function $\mathcal{F}(\gamma)$ can be considered as a complex injection function composed of two terms: a scaled steady-state electron spectrum and a scaled injection spectrum. From Eq.~\ref{eq:initcond1} and \ref{eq:decompose} we deduce that the initial condition for $N_{\text{e,p}}(\gamma,t)$:

\begin{equation} \label{eq:initcondpert}
    N_{\text{e,p}}(\gamma,t=0) = 0
\end{equation}

\subsection{Solving by characteristics}

We use the method of characteristics to solve the Eq.~\ref{eq:kineqforperturb}. We first search for characteristic curves in the $\gamma$--$t$ space along which the equation for $N_{\text{e,p}}(\gamma,t)$ becomes an ordinary differential equation. Then we solve this equation along a characteristic curve. Let us rewrite the Eq.~\ref{eq:kineqforperturb} in the following form (expanding the partial derivative over $\gamma$):

\begin{multline*} 
 		\dfrac{\partial N_{\text{e,p}}(\gamma,t)}{\partial t} \, + \,  (-1) \, W(\gamma) \, \dfrac{\partial N_{\text{e,p}}(\gamma,t)}{\partial \gamma} = \mathcal{F}(\gamma) - \dfrac{N_{\text{e,p}}(\gamma,t)}{\tau(\gamma)} 
\end{multline*}

\vspace*{2mm}

where $\frac{1}{\tau(\gamma)} = \frac{1}{t_{\text{esc}}} + \frac{1}{t_{_{\text{FI}}}} - 2 b_\text{c} \gamma$. \\

Let us consider a characteristic curve ($\gamma(t),t$). The left hand side of the equation can be now represented as a full derivative of $N_{\text{e,p}}(\gamma(t),t)$ with respect to time, and also as a directional derivative of $N_{\text{e,p}}(\gamma(t),t)$ in the direction of ($-W(\gamma) \, , \, 1$) in the $\gamma$--$t$ plane. By the chain rule, we have:

\begin{equation*}
    \dfrac{dN_{\text{e,p}}(\gamma(t),t)}{dt} = \dfrac{\partial N_{\text{e,p}}(\gamma(t),t)}{\partial t} + \dfrac{d\gamma(t)}{dt} \cdot \dfrac{\partial N_{\text{e,p}}(\gamma(t),t)}{\partial \gamma}
\end{equation*}

We see that along the characteristic curve ($\gamma(t),t$) our equation in partial derivatives transforms into an ordinary differential equation:

\begin{equation} \label{eq:fullderivt}
 		 \dfrac{dN_{\text{e,p}}(\gamma(t),t)}{dt} = \mathcal{F}(\gamma) - \dfrac{N_{\text{e,p}}(\gamma(t) \, ,t)}{\tau(\gamma)}
 \end{equation} 
 	
 \begin{equation} \label{eq:dgammadt}
 	\dfrac{d\gamma(t)}{dt} = - W(\gamma)
 	\end{equation}

Let us solve the Eq.~\ref{eq:dgammadt} for the characteristic curve in the $\gamma$--$t$ space. We choose an initial point on our characteristic as ($\xi$,0), so the equation has to satisfy the boundary condition $\gamma(t=0) = \xi$. The solution of the Eq.~\ref{eq:dgammadt} with this boundary condition is:

\begin{equation} \label{eq:gammaxit}
 		 \gamma_{(\xi)}(t) = \dfrac{1}{b_\text{c} t_{_{\text{FI}}} (1 - e^{-t/t_{_{\text{FI}}}}) + \frac{1}{\xi} e^{-t/t_{_{\text{FI}}}}}
 \end{equation}  		

This formula defines a characteristic curve in the $\gamma$-$t$ space. For given $\gamma$ and $t$, let us find the starting Lorentz factor $\xi$ of the characteristic that passes through point ($\gamma$,$t$):

\begin{equation} \label{eq:xigammat}
    \xi = \xi(\gamma,t) = \dfrac{\gamma \, e^{-t/t_{_{\text{FI}}}}}{1 - b_\text{c} t_{_{\text{FI}}} \gamma \, (1 - e^{-t/t_{_{\text{FI}}}})}
\end{equation}

Now let us solve the initial value problem (Eq.~\ref{eq:fullderivt} and \ref{eq:initcondpert}). We restrict the $N_{\text{e,p}}(\gamma,t)$ to the characteristic (Eq.~\ref{eq:gammaxit}), noting $N_{\text{e,p}}(\gamma_{(\xi)}(t),t)$ $\Rightarrow$ $u(t)$ at a given $\xi$ and solve the differential equation Eq.~\ref{eq:fullderivt} along the characteristic curve. We have:

\begin{equation*}
 		 \dfrac{du(t)}{dt} +  \dfrac{u(t)}{\tau(\gamma_{(\xi)}(t))} = \mathcal{F}(\gamma_{(\xi)}(t))
 \end{equation*} 

This is a simple linear non-homogeneous first order differential equation, which can be solved with the help of an integrating factor. The equation has the following general solution:

\begin{equation} \label{eq:gensoldefinint}
    u(t) = \dfrac{1}{\mu(t)} \left[ \, \int_{0}^{t} \mu(t^{\prime}) \, \mathcal{F}(\gamma_{(\xi)}(t^{\prime})) \, dt^{\prime} \, + \, C \, \right]
\end{equation}

\vspace*{2mm}

with the integrating factor:

\begin{equation} \label{eq:intfactexpr}
    \mu(t) = e^{\int dt / \tau(\gamma_{(\xi)}(t))} 
\end{equation}

From the initial condition Eq.~\ref{eq:initcondpert} which is $u(t=0) = 0$, we get the constant of integration $C = 0$.

Now let us calculate the $\mu(t)$ function. First we evaluate the exponent in Eq.~\ref{eq:intfactexpr}:

\begin{multline*}
    \int \frac{dt}{\tau(\gamma_{(\xi)}(t))} \, = \, \int \, \left(\frac{1}{t_{\text{esc}}} + \frac{1}{t_{_{\text{FI}}}}\right) \, dt \, - \\ - \, 2 b_\text{c} \int \dfrac{1}{b_\text{c} t_{_{\text{FI}}} + (1/\xi - b_\text{c} t_{_{\text{FI}}}) \, e^{-t/t_{_{\text{FI}}}}}  \, dt \, = \\ = \, \left[\frac{1}{t_{\text{esc}}} + \frac{1}{t_{_{\text{FI}}}}\right] \, t \, - \, 2 \, \text{ln}\left[1/\xi \, + \, b_\text{c} \, t_{_{\text{FI}}} \, (e^{t/t_{_{\text{FI}}}} - 1)\right]
\end{multline*}

\vspace*{7mm}

The integrating factor $\mu(t)$ is then:

\begin{equation} \label{eq:intfactxit}
    \mu(t) = \dfrac{e^{(1/t_{\text{esc}} \, + \, 1/t_{_{\text{FI}}}) \, t}}{\left[1/\xi \, + \, b_\text{c} t_{_{\text{FI}}} (e^{t/t_{_{\text{FI}}}} - 1)\right]^2}
\end{equation}

\subsection{Final solution}

The transition from $u(t)$ back to $N_{\text{e,p}}(\gamma,t)$ is achieved by substitution of $\xi=\xi(\gamma,t)$ to the expression for $u(t)$ (Eq.~\ref{eq:gensoldefinint}): $N_{\text{e,p}}(\gamma,t) = u(t)|_{\xi=\xi(\gamma,t)}$.  

\begin{equation} \label{eq:finalsolgen}
    N_{\text{e,p}}(\gamma,t) = u(t)|_{\xi=\xi(\gamma,t)} = \int_{0}^{t} \left[ \dfrac{\mu(t^{\prime})}{\mu(t)} \, \mathcal{F}(\gamma_{(\xi)}(t^{\prime})) \right]_{|_{\xi=\xi(\gamma,t)}} \, dt^{\prime}
\end{equation}

Let us proceed with the substitution $\xi=\xi(\gamma,t)$ to the components of the integrand. 

First we evaluate $\mu(t)|_{\xi=\xi(\gamma,t)}$. Substituting the expression for the initial Lorentz factor $\xi = \xi(\gamma,t)$ from Eq.~\ref{eq:xigammat}, we get:

\begin{equation} \label{eq:mutgamma}
    \mu(t)|_{\xi=\xi(\gamma,t)} = \mu(t,\gamma) = \gamma^2 \, e^{(1/t_{\text{esc}} \, - \, 1/t_{_{\text{FI}}}) \, t}
\end{equation}

Next, we compute the form of the $\mu(t^{\prime})|_{\xi=\xi(\gamma,t)}$, again substituting the expression for $\xi$ from Eq.~\ref{eq:xigammat}:

\begin{equation} \label{eq:mutprimegammat}
    \mu(t^{\prime})|_{\xi=\xi(\gamma,t)} \, = \, \mu(t^{\prime},\gamma,t) \, = \, \dfrac{e^{t^{\prime}/t_{\text{esc}} \, + \, (t^{\prime} - 2t)/t_{_{\text{FI}}}}}{\left[1/\gamma \, + \, b_\text{c} \, t_{_{\text{FI}}} \, (e^{(t^{\prime} - t)/t_{_{\text{FI}}}} - 1)\right]^2}
\end{equation}

Then we have to calculate the Lorentz factor $\gamma_{(\xi)}(t^{\prime})|_{\xi=\xi(\gamma,t)}$ that appears in the function $\mathcal{F}(\gamma_{(\xi)}(t^{\prime}))$. We use Eq.~\ref{eq:gammaxit}, \ref{eq:xigammat} and after simple and obvious transformations we obtain:

\begin{equation} \label{eq:biggamma}
 \gamma_{(\xi)}(t^{\prime})|_{\xi=\xi(\gamma,t)} = \Gamma(\gamma,t,t^{\prime}) = \dfrac{\gamma \, e^{(t^{\prime} - t)/t_{_{\text{FI}}}}}{1 \, + \, \gamma b_\text{c} \, t_{_{\text{FI}}} \, (e^{(t^{\prime} - t)/t_{_{\text{FI}}}} \, - \, 1)}
\end{equation}

We note that the denominator of $\mu(t^{\prime},\gamma,t)$ in Eq.~\ref{eq:mutprimegammat} multiplied by $\gamma^2$ is exactly the square of denominator of the $\Gamma(\gamma,t,t^{\prime})$, so for simplicity we express $\mu(t^{\prime},\gamma,t)$ via $\Gamma(\gamma,t,t^{\prime})$:

\begin{equation} \label{eq:mutprimesimpl}
   \mu(t^{\prime},\gamma,t) = \Gamma^2(\gamma,t,t^{\prime}) \, \, e^{(1/t_{\text{esc}} \, - \, 1/t_{_{\text{FI}}}) \, t^{\prime}} 
\end{equation}

Now we evaluate the expression under the integral in Eq.~\ref{eq:finalsolgen}, using previously derived components, where the substitution was done (Eq.~\ref{eq:mutgamma}, \ref{eq:mutprimesimpl} and \ref{eq:biggamma}):

\begin{multline} \label{eq:integrandnp}
 \left[ \dfrac{\mu(t^{\prime})}{\mu(t)} \, \mathcal{F}(\gamma_{(\xi)}(t^{\prime})) \right]_{|_{\xi=\xi(\gamma,t)}} = \dfrac{\Gamma^2(\gamma,t,t^{\prime})}{\gamma^2} \, e^{(1/t_{\text{esc}} \, - \, 1/t_{_{\text{FI}}}) \, (t^{\prime} \, - \, t)} \, \cdot \\ \cdot \, \left[ \dfrac{N_{\text{e},0}(\Gamma(\gamma,t,t^{\prime}))}{t_{_{\text{FI}}}} \, \left( 1 - \frac{1}{b_\text{c} t_{\text{esc}} \Gamma(\gamma,t,t^{\prime})} \right) \, + \, \dfrac{Q_{\text{inj}}(\Gamma(\gamma,t,t^{\prime}))}{b_\text{c} t_{_{\text{FI}}} \Gamma(\gamma,t,t^{\prime})} \right] = \\ = \dfrac{\Gamma(\gamma,t,t^{\prime}) \, e^{(1/t_{\text{esc}} \, - \, 1/t_{_{\text{FI}}}) \, (t^{\prime} \, - \, t)}}{b_\text{c} \, t_{_{\text{FI}}} \, \gamma^2} \, \cdot \\ \cdot \, \left[ Q_{\text{inj}}(\Gamma(\gamma,t,t^{\prime})) \, + \, \left(b_\text{c} \Gamma(\gamma,t,t^{\prime}) \, - \, \frac{1}{t_{\text{esc}}} \right) \, N_{\text{e},0}(\Gamma(\gamma,t,t^{\prime})) \right] 
\end{multline}

We can now write down the final solution for the total electron spectrum $N_{\text{e,FI}}(\gamma,t)$ using Eq.~\ref{eq:decompose}, \ref{eq:finalsolgen} and \ref{eq:integrandnp}:

\begin{multline} \label{eq:finalsolutneshock}
    N_{\text{e,FI}}(\gamma,t) = N_{\text{e},0}(\gamma) \, + \, \int_{0}^{t} \dfrac{\Gamma(\gamma,t,t^{\prime}) \cdot \text{exp} \! \left[(1/t_{\text{esc}} \, - \, 1/t_{_{\text{FI}}}) \, (t^{\prime} \, - \, t)\right]}{b_\text{c} \, t_{_{\text{FI}}} \, \gamma^2} \, \cdot \\ \cdot \, \left[ Q_{\text{inj}}(\Gamma(\gamma,t,t^{\prime})) \, + \, \left(b_\text{c} \Gamma(\gamma,t,t^{\prime}) \, - \, 1 / t_{\text{esc}} \right) \, N_{\text{e},0}(\Gamma(\gamma,t,t^{\prime})) \right]  \, dt^{\prime}
\end{multline}

Let us explore the final solution. At the moment when the shock just enters the blob ($t=0$), the electron spectrum is, as expected, the steady-state solution. Also, when the shock acceleration is extremely weak ($t_{_{\text{FI}}} \rightarrow \infty$), we see that the electron spectrum will remain the steady-state one and not evolve in time, which is in agreement with the expectations (very weak shock will not perturb the electron spectrum).

%%%%%%%%%%%%%%%%%%%%%%%%%%%%%%%%%%%%%%%%%%%%%%%%%%

% Don't change these lines
\bsp	% typesetting comment
\label{lastpage}
\end{document}